\documentclass[12pt]{article}
\usepackage[dvipdfmx]{graphicx}
\usepackage{amsmath}
\usepackage{amssymb}
\usepackage{authblk}
\usepackage{mathrsfs}
\usepackage{slashed}
\usepackage{color}
\usepackage{cite}

\makeatletter
 
 \@addtoreset{equation}{section}
 \makeatother
\newcommand{\be}{\begin{equation}}
\newcommand{\ee}{\end{equation}}
\newcommand{\bea}{\begin{eqnarray}}
\newcommand{\eea}{\end{eqnarray}}
\newcommand{\beann}{\begin{eqnarray*}}
\newcommand{\eeann}{\end{eqnarray*}}
\newcommand{\nn}{\nonumber}
\newcommand{\ba}{\begin{array}}
\newcommand{\ea}{\end{array}}

\newcommand{\bs}{\boldsymbol}
\newcommand{\D}{\mathcal{D}}

\newcommand{\Phib}{\bar{\Phi}}
\newcommand{\Qt}{\tilde{Q}}
\newcommand{\del}{\partial}

\DeclareMathOperator{\im}{im}

\DeclareMathOperator{\rank}{rank}
\DeclareMathOperator{\diag}{diag}
\DeclareMathOperator{\ind}{ind}
\DeclareMathOperator{\Tr}{Tr}
\DeclareMathOperator{\Spec}{Spec}

\title{Supersymmetric Gauge Theory on the Graph}
\author[1]{So Matsuura\thanks{s.matsu@phys-h.keio.ac.jp}}
\author[2]{Kazutoshi Ohta\thanks{kohta@law.meijigakuin.ac.jp}}
\affil[1]{\it Hiyoshi Departments of Physics, 
and Research and Education Center for Natural Sciences,
Keio University, 4-1-1 Hiyoshi, Yokohama, Kanagawa 223-8521, Japan}
\affil[2]{\it Institute of Physics, Meiji Gakuin University, Yokohama, Kanagawa 244-8539, Japan}

\date{}

\begin{document}
\maketitle


\begin{center}
{\bf Abstract}
\end{center}

We consider two-dimensional ${\cal N}{=}(2,2)$ supersymmetric gauge theory
on discretized Riemann surfaces.
We find 
that the discretized theory can be efficiently described by using graph theory, where
the bosonic and fermionic fields are regarded
as vectors on a graph and its dual. We first analyze the Abelian theory and identify its spectrum in terms of graph theory. 
In particular, we show that the fermions have zero modes corresponding to the topology of the graph, 
which can be understood as kernels of the incidence matrices 
of the graph and the dual graph. 
In the continuous theory, 
a scalar curvature appears as an anomaly in the Ward-Takahashi (WT) identity associated with a $U(1)$ symmetry. 
We find that the same anomaly arises
as the deficit angle at each vertex on the graph. 
By using the localization method,
we show that the path integral on the graph reduces to 
an integral over a set of the zero modes. 
The partition function is then ill-defined unless suitable operators are inserted. 
We extend the same argument to the non-Abelian theory 
and show that the path integral reduces to multiple integrals of 
Abelian theories at the localization fixed points.

\newpage

\section{Introduction}

Gauge theory has been recognized as the most fundamental theory describing the interaction of elementary particles.
However, in recent years, the importance of gauge theories has been extended and
it has become to be recognized as a powerful tool for approaching the quantum theory of gravity through the gauge/gravity duality \cite{Maldacena:1997re,Witten:1998qj,Gubser:1998bc}. 
In particular, the gauge/gravity duality predicted by the superstring theory provides a clear dictionary between the gauge and gravity theories with supersymmetry.
Since supersymmetric gauge theories can give mathematically accurate descriptions for some physical quantities by using strong constraints based on supersymmetry, they have traditionally been studied in analytic ways.
However, the prediction of the gauge/gravity duality should also be applied to the dynamical quantities. 
To approach quantum gravity through the supersymmetric gauge theories, 
therefore, we need to have the means to analyze their dynamics in a non-perturbative way. 

One of the most effective non-perturbative approaches to 
gauge theories is lattice gauge theory, 
which regularizes gauge field theory on a finite lattice 
and defines the continuous theory as the continuum limit of the discretized theory. 
Various attempts have been made to construct the lattice gauge theories 
with preserving a part of supersymmetries on the square lattice
\cite{
Sugino:2003yb,
Sugino:2004qd,
Sugino:2004uv,
Sugino:2006uf,
Sugino:2008yp,
Kikukawa:2008xw,
Kaplan:2002wv,
Cohen:2003xe,
Cohen:2003qw,
Kaplan:2005ta,
DAdda:2004dmn,
DAdda:2005rcd,
DAdda:2007hnx,
Endres:2006ic,
Giedt:2006dd,
Matsuura:2008cfa,
Catterall:2003wd,
Joseph:2013jya,
Catterall:2007kn}.
The relations between these models were investigated in \cite{Unsal:2005yh,Unsal:2006qp,Takimi:2007nn,Damgaard:2007xi,Damgaard:2007eh}, 
and several numerical computations based on these models 
have been carried out 
\cite{Catterall:2006jw,
Suzuki:2007jt,
Kanamori:2007ye,
Kanamori:2007yx,
Kanamori:2008bk,
Kanamori:2008yy,
Kanamori:2009dk,
Hanada:2009hq,
Kadoh:2009rw,
Catterall:2008dv,
Catterall:2011aa,
Catterall:2012yq,
Catterall:2014vka,
Giguere:2015cga,
Catterall:2017xox}.
In two dimensions, 
the theory is super renormalizable, 
and numerical calculations can be performed without maintaining supersymmetry with a small number of fine-tunings
\cite{Suzuki:2007jt}.
In \cite{August:2018esp}, 
two-dimensional ${\cal N}=(2,2)$ supersymmetric $SU(2)$ Yang-Mills theory is numerically investigated 
by using a lattice theory with a conventional Wilson fermion. 
For reviews, see \cite{Kaplan:2003uh,Giedt:2006pd,Catterall:2009it,Joseph:2011xy,Sugino:2013mqa,Bergner:2016sbv}.

Among these lattice theories,  
the models constructed by Sugino (Sugino models) 
\cite{Sugino:2003yb,Sugino:2004qd,Sugino:2004uv,Sugino:2006uf,Sugino:2008yp}
have the gauge group $SU(N)$, 
while the gauge group of the other models is inevitably $U(N)$. 
As a result, the link variables in the Sugino models are 
expressed by compact unitary matrices 
as in the conventional lattice gauge theories. 
In the Sugino models, 
the action is written by an exact form of scalar supercharges constructed by topological twisting.
Then these scalar supersymmetries are manifestly preserved 
even if the translational symmetry is explicitly broken 
by the discretization of space-time. 
The problem of vacuum degeneracy of lattice gauge field 
has been solved without using an admissibility condition \cite{Matsuura:2014pua} 
and the tree level improvement has been proposed 
\cite{Hanada:2017gqc}.

In \cite{Matsuura:2014kha}, the Sugino model defined on the usual square lattice is extended to a theory on a discrete space-time where the two-dimensional Riemann surface is divided by polygons. This model (the generalized Sugino model) has been subjected to rigorous analysis using the method of localization \cite{Matsuura:2014nga} and numerical calculations \cite{Kamata:2016xmu}.

In this paper, we reconstruct and analyze the generalized Sugino model in two-dimensional supersymmetric gauge theories by using graph theory.
(For an introduction to graph theory, see e.g.~\cite{bapat2010graphs}.) 
We regard the vertices and edges of the polygons on the discretized Riemann surface as a graph. 
We assume that the graph is a directed graph with an edge orientation.  
We also assume that faces are assigned to the vertices of the dual graph.
The directed graph can introduce a ``difference'' between adjacent vertices, and the difference operator expressed by a matrix is called an incidence matrix. Using this incidence matrix, it is possible to construct a field theory on the discrete space-time represented by the graph, including supersymmetry.

The generalized Sugino model has been analyzed using the method of localization owing to supersymmetry in \cite{Matsuura:2014nga}, 
but the use of graph theory makes it possible to discuss clearer.
In particular, 
we can see the structure of zero modes, 
which play an important role in the localization.  
The zero modes appear as a kernel of the incidence matrix in the context of graph theory,
and thus we can use linear algebra to understand their property.
In addition, 
the analysis by graph theory is beneficial in understanding the anomaly as well 
since the zero modes are important in understanding the anomaly of the theory 
even in the discretized theory. 
For analysis of field theories and quantum mechanics on the graph
including the supersymmetry from other viewpoints, see \cite{Kan:2004kz,Kan:2009tu,Kan:2013mra}.
See also \cite{Ohta:2020ygi} for localization in the quiver gauge theory by using 
the technology of graph theory.

The organization of this paper is as follows. 
In Sec.~\ref{SUSY on Riemann},
we briefly review the construction of
${\cal N}{=}(2,2)$ supersymmetric Yang-Mills theory
on the smooth Riemann surface
by using differential forms.
The formulation by the differential forms
makes the relation to the graph structure clearer later.
We also derive the currents and the Ward-Takahashi(WT) identities corresponding to the global symmetry of the theory. 
In Sec.~\ref{Graph Theory for SUSY YM},
we prepare some basics of graph theory for the discretization of the
Riemann surface. We also introduce some useful matrices including the incidence matrix
and summarize their properties in graph theory.
In Sec.~\ref{Abelian Theory},
we formulate a supersymmetric Abelian gauge theory
by using graph theory and
discuss the properties of the fermion zero modes.
We derive the chiral anomaly on the graph
and show that the fermion zero modes play an important role. 
We see that the anomaly in the WT identity,
which appears as the scalar curvature in the continuous theory, 
appears as the deficit angle in the theory on the graph.
In Sec.~\ref{Localization in the Abelian theory},
we perform the path integral for the Abelian theory by
using the localization method.
We see that there exists a residual integral over the zero modes
after integrating out non-zero modes.
The integral over the zero modes makes the partition function itself ill-defined,
but we also discuss a remedy for this problem by inserting operators
including bi-linear terms of the fermions.
In Sec.~\ref{Non-Abelian Theory},
we generalize the localization arguments to non-Abelian theory.
By the saddle point approximation in the localization method,
non-zero modes give a Vandermonde type measure and
the path integral reduces to multiple integrals of
the Abelian theory one. For this non-Abelian theory,
we will see the zero modes play an important role as well as the Abelian theory.
Sec.~\ref{Conclusion and Discussion} is devoted to 
conclusion and discussion. 
In Appendix, we give some concrete examples of the graph structure and
properties of the incidence matrix and Laplacian. We also give the
convention of Weyl-Cartan bases, which are used for non-Abelian theory.

\section{Supersymmetric Gauge Theory on the Riemann Surface}
\label{SUSY on Riemann}

\subsection{Action and currents in supersymmetric gauge theory}

We start with the review of the supersymmetric gauge theory on the smooth Riemann surface $\Sigma_h$
(continuous space-time), which has the genus (handles) $h$.
The theory considering in this paper is essentially obtained by  dimensional reduction from four-dimensional ${\cal N}{=}1$ supersymmetric gauge theory with four supercharges, namely two-dimensional ${\cal N}{=}(2,2)$ supersymmetric gauge theory. 
In general, however, when one simply constructs this theory on a curved manifold, the supersymmetry is completely broken. 
The point is that a part of supersymmetry can be restored by introducing a specific $U(1)$ gauge field as a background in accordance with the spin connection of the background space-time. 
Supersymmetric theory obtained by this procedure becomes naturally a topologically twisted theory on the curved space-time, and a half of the supersymmetry is recovered in general. 
Among the ways of the topological twistings, which depend on how to turn on the background gauge field, we choose the so-called topological A-model through, 
which is the theory that we consider in this paper.
(See also \cite{Ohta:2019odi,Ohta:2020ygi} for more detailed construction.)

Because of the topological twisting, not only the bosonic fields 
but also the fermion fields have integer spins. 
Therefore it is convenient to express the fields in this theory by differential forms: 
The 0-form scalar fields are $\Phi$, $\Phib$ and $\eta$, 
the 1-form vector fields are $A\equiv A_\mu dx^\mu$ and $\lambda\equiv \lambda_\mu dx^\mu$,
and 2-form fields are $Y\equiv \frac{1}{2}Y_{\mu\nu}dx^\mu \wedge dx^\nu$
and
$\chi\equiv \frac{1}{2}\chi_{\mu\nu}dx^\mu \wedge dx^\nu$, 
where $\Phi$, $\Phib$, $A$ and $Y$ are bosons and $\eta$, $\lambda$ and $\chi$ are
fermions (Grassmann valued). 

We write $Q$ for one of the supercharges, which transforms the fields as%
\footnote{The notation of the $Q$ transformation has been slightly changed from the previous works \cite{Matsuura:2014kha,Matsuura:2014nga,Kamata:2016xmu}.}
\begin{equation}
\begin{array}{lcl}
Q \Phi = 0, && \\
Q \bar{\Phi} = 2\eta, &&  Q \eta = \frac{i}{2}[\Phi,\bar{\Phi}],\\
Q A = \lambda, && Q \lambda =  -d_A \Phi,\\
Q Y = i[\Phi,\chi], && Q\chi = Y,
\end{array}
\end{equation}
where $d_A \Phi \equiv d\Phi + i[A,\Phi]$ is a covariant exterior derivative for the  adjoint scalar field.
We can see that the square of $Q$ generates the gauge transformation with a parameter $\Phi$, which is denoted by
$Q^2 = \delta_\Phi$.

Using this supercharge, we can write the action of theory in $Q$-exact form
\be
S
= -\frac{1}{2g^2}Q \int_{\Sigma_h} \Tr \bigg\{
\frac{i}{2}\eta[\Phi,\bar{\Phi}]\omega
+d_A\bar{\Phi} \wedge {*} \lambda 
+\chi {*} (Y-2F)
\bigg\},
\label{Q-exact action}
\ee
where $\omega$ is a volume (K\"ahler) form on $\Sigma_h$, $*$ represents the Hodge star operation
which maps from an $n$-form to a $(2-n)$-form, and 
\be
F \equiv dA+iA\wedge A
\ee
is a field strength, which will give a kinetic term of the gauge field after integrating out
the auxiliary filed $Y$.

More concretely, after applying the $Q$-transformation, the bosonic and fermionic parts of the action are given by
\bea
S_B &=& \frac{1}{2g^2}  \int_{\Sigma_h} \Tr\bigg\{
\frac{1}{4}[\Phi,\bar{\Phi}]^2 \omega
+ d_A\Phib \wedge {*} d_A \Phi
-Y {*} (Y-2F)
\bigg\},\\
S_F &=& \frac{1}{2g^2}  \int_{\Sigma_h} \Tr\bigg\{
i \eta[\Phi,\eta]\omega
+2\eta  d_A {*}\lambda
-i\lambda \wedge{*}[\Phib,\lambda]
+i\chi {*}[\Phi,\chi]
- 2\chi {*}d_A\lambda
\bigg\},
\label{fermionic action}
\eea
respectively.

For later convenience, we define a vector of fermionic fields in the following order,
\be
\Psi \equiv
\begin{pmatrix}
\eta\\
\chi\\
\lambda
\end{pmatrix}. 
\ee
Then the fermionic part of the action (\ref{fermionic action}) reduces to
\be
S_F = \frac{1}{2g^2} \Tr \int_{\Sigma_h}
\Psi^T \wedge {*}\left(i{\slashed{\cal D}}_A+{\cal M}_\Phi\right) \Psi,
\ee
where $\slashed{\cal D}_A$ and ${\cal M}_\Phi$ are 
the Dirac operator and the mass matrix depending on $\Phi$ and $\Phib$, respectively; 
\be
\slashed{\cal D}_A
\equiv
\begin{pmatrix}
0 & 0 &  id_A^\dag \\
0 & 0 & id_A \\
-i d_A & -i d_A^\dag & 0
\end{pmatrix},
\quad
{\cal M}_\Phi
\equiv
\begin{pmatrix}
i[\Phi,\cdot] & 0 & 0\\
0 & i[\Phib,\cdot] & 0\\
0 & 0 & -i[\Phi,\cdot] \\
\end{pmatrix}.
\label{Dirac on Sigma}
\ee
Here 
\be
d_A^\dag \equiv -{*}d_A{*}
\ee
is the co-differential operator, which maps from an $n$-form to an $(n-1)$-form
on $\Sigma_h$ and $[\Phi,\cdot]$ represents an adjoint action induced by $\Phi$.


Let us now consider yet another supercharge $\Qt$.
Since there are two preserved supercharges on the curved Riemann surface, we have the supercharge $\Qt$
in addition to $Q$.
The supersymmetry transformation for the vector multiplet is given by
\be
\begin{array}{lcl}
\Qt \Phi = 0, && \\
\Qt A = {*}\lambda, && \Qt \lambda =  {*}d_A \Phi,\\
\Qt \bar{\Phi} = 2{*}\chi, &&  \Qt \chi = \frac{i}{2}[\Phi,\bar{\Phi}]\omega,\\
\Qt Y = -i[\Phi,\eta]\omega, && \Qt \eta = -{*}Y.
\end{array}
\ee
Roughly speaking, $\Qt$ swaps the role of 0-form $\eta$ and 2-form $\chi$ against the action of $Q$.
We can also see the square of $\Qt$ becomes the gauge transformation, namely
$\Qt^2 = \delta_\Phi$ and $Q$ and $\Qt$ is anti-commuting with each other; 
\be
\{Q,\Qt\}=0.
\ee

Using the transformation of $\Qt$, we can also write the action in the $\Qt$-exact form
\be
S =-\frac{1}{2g^2}\Qt  \int_{\Sigma_h}
\Tr\Bigg\{
\frac{i}{2}\chi[\Phi,\bar{\Phi}]
+\lambda \wedge d_A\bar{\Phi}
-\eta (Y-2F)
\Bigg\}.
\ee
So the action is also invariant under $\Qt$. This obeys from the fact that the action
is written by
\be
S = \frac{1}{4g^2}[Q,\Qt]\int_{\Sigma_h}  \Tr\left\{
\Phib F + \eta \chi
\right\}.
\label{QQt exact form}
\ee
Using also the anti-commuting relation between $Q$ and $\Qt$, we can find that the action is written 
in both $Q$- and $\Qt$-exact forms.
Note also that the part acting the supercharges in (\ref{QQt exact form})
is invariant under swapping (a rotation of) $\eta$ and $\chi$.

\subsection{Symmetries and relations among conserved currents}

We next consider the global symmetries of this theory and the associated Noether currents. 
In general, if a theory is invariant under a certain global transformation, 
the infinitesimal transformation of the action can be written as
\begin{equation}
  \delta_\xi S = \xi\,\bs{s} S = \int_{\Sigma_h}\left( d\xi * \tilde{J}_s + \xi dI_s \right), 
\end{equation}
where $\bs{s}$ is the generator of this transformation, 
$\xi$ is a position-dependent parameter, 
and $\tilde{J}_s$ and $I_s$ are both one-form.
Then the corresponding Noether current is defined by
\begin{equation}
  J_s = \tilde{J}_s + *I_s, 
\end{equation}
which is conserved at least classically; 
\begin{equation}
  d^\dagger J_s = 0.
\end{equation}
Note that this current is invariant if we add any $s$-invariant total derivative term to the action. 

Using this prescription, the Noether current corresponding to $Q$ and $\tilde{Q}$ symmetries are constructed as 
\be
J_Q = \frac{1}{g^2}\Tr\left\{
d_A\Phi \eta
+{*}d_A\Phi {*}\chi
-\frac{i}{2}[\Phi,\Phib]\lambda
-{*}Y {*}\lambda
\right\}, 
\ee
and 
\be
J_{\tilde{Q}}  = \frac{1}{g^2}\Tr\left\{
-{*}d_A\Phi \eta
+d_A\Phi {*}\chi
-\frac{i}{2}[\Phi,\Phib]{*}\lambda
+{*}Y \lambda
\right\}, 
\ee
respectively. 

The theory also possesses two global $U(1)$ symmetries, $U(1)_A$ and $U(1)_V$.
The $U(1)_A$ symmetry transforms the fields as 
\be
\delta_A A =0, \quad 
\delta_A \Phi = 2i\theta_A \Phi,\quad
\delta_A\Phib=-2i\theta_A\Phib,\quad
\delta\Psi = i\theta_A \gamma_A \Psi,
\ee
where
\be
\gamma_A
\equiv 
\begin{pmatrix}
-1 & 0 & 0 \\
0 & -1 & 0 \\
0 & 0 & 1
\end{pmatrix}\,,
\ee
and the $U(1)_V$ symmetry transforms the fields as 
\be
\delta_V A =0, \quad 
\delta_V \Phi = 0, \quad  
\delta_V \bar\Phi = 0, \quad 
\delta_V \Psi = i\theta_V \gamma_V \Psi,
\ee
where 
\be
\gamma_V \equiv
-i\begin{pmatrix}
0 & {*} & 0\\
-* & 0 & 0\\
0 & 0 & {*}
\end{pmatrix}.
\ee
The corresponding Noether currents are given by
\begin{align}
  J_A &= \frac{i}{g^2} {\rm Tr}\left(
    -\bar\Phi d_A \Phi + \Phi d_A \bar\Phi + \eta \lambda - * \lambda * \chi
  \right)\,, \\
  J_V &= \frac{1}{g^2} {\rm Tr}\left( \eta * \lambda + \lambda * \chi \right)\,. 
\end{align}
Note that $\gamma_A$ and $\gamma_V$ satisfy $\gamma_A^2 = \gamma_V^2 = 1$.
We will soon see that the $U(1)_A$ symmetry is anomalous quantum mechanically.

We point out that there are important relationships among the supercurrents $J_Q$, $J_{\tilde{Q}}$ and the current $J_V$ \cite{Kadoh:2009rw}; 
\bea
QJ_V &=& -J_{\tilde{Q}},\\
\tilde{Q}J_V &=& J_Q.
\eea
This means that 
the conservation law of the $U(1)_V$ symmetry 
guarantees the conservation law of the $\tilde{Q}$ symmetry; 
\be
d^\dagger J_V=0 \quad \Rightarrow \quad d^\dagger J_{\tilde{Q}}=0,
\ee
if the $Q$ symmetry is preserved.

\subsection{Ward-Takahashi identities and anomaly}
\label{sec:WT continuum} 

We next consider the WT identities.
In the path-integral formalism, 
the WT identity is derived from the obvious invariance 
under a change of variables, 
\begin{equation}
  \int dX \, {\cal O}(x_0) e^{-S[X]} = \int dX' \, {\cal O}'(x_0) e^{-S[X']}\,,
  \label{eq:basic id}
\end{equation}
where $X$ expresses the fields of the theory,
$X'$ is transformed fields of $X$,%
\footnote{We assume that the ranges of the integration by $X$ and $X'$ are identical.}
and ${\cal O}(x_0)$ is a local operator at a position $x_0$. 
Note here that this identity can be applied regardless of whether the classical action
is invariant under the transformation or not. 
So if we consider a general transformation with the parameter $\xi(x)$ and the generator $\bs{s}$,
the infinitesimal transformation of the action $S$ and the operator ${\cal O}$ are given by
\begin{equation}
  \delta S = \int_{\Sigma_h} dx \, \xi(x) K_s(x)\,,
\end{equation}
and 
\begin{equation}
  \delta {\cal O}(x_0) = \xi(x_0)\, \bs{s}{\cal O}(x_0)\,,
\end{equation}
respectively.
In addition, we assume that the integration measure transforms as
\begin{equation}
  dX' = dX \left(1 + \int_{\Sigma_h} dx\, \xi(x) {\cal A}_s(x)\right)\,.
\end{equation}
Then, from \eqref{eq:basic id}, we obtain the identity for the vacuum expectation value (vev), 
\begin{equation}
    \int_{\Sigma_h} dx \, \xi(x) \Bigl\langle \bs{s}{\cal O}(x_0)\delta(x-x_0) 
    - {\cal O}(x_0) \left( K_s(x) - {\cal A}_s(x)\right) \Bigr\rangle = 0\,,
\end{equation}
where the vev is defined by
\begin{equation}
  \left\langle {\cal O}(x) \right\rangle \equiv {Z}^{-1}\int dX\, {\cal O}(x) e^{-S[X]}\,,
\end{equation}
with the partition function $Z=\int dX \, e^{-S[X]}$. 
In particular, if we assume the parameter $\xi(y)$ takes the form $\xi(y)=\xi \delta(y-x)$ for a specific coordinate $x$ 
and constant parameter $\xi$, 
we obtain the identity for the vev of local variables;
\begin{equation}
   \Bigl\langle {\cal O}(x_0) \left( K_s(x) - {\cal A}_s(x)\right) \Bigr\rangle 
  =\delta(x-x_0) 
  \Bigl\langle \bs{s}{\cal O}(x_0) \Bigr\rangle  \,. 
\end{equation}

Let us return to the supersymmetric gauge theory that we are considering. 
As discussed in the previous subsection, the action of the supersymmetric theory
is invariant under the two supersymmetry transformations generated by $Q$ and $\tilde{Q}$, and the two $U(1)$ transformations $U(1)_A$ and $U(1)_V$. Therefore, the transformation of the action \eqref{Q-exact action} is given by 
\begin{equation}
  \delta_s S = \int_{\Sigma_h} dx \, \xi(x) d^\dagger J_s(x) \omega \,,
\end{equation}
up to total derivative, where $J_s(x)$ is the corresponding Noether current.

Although it is sufficient only to consider the action \eqref{Q-exact action} 
of the continuous theory, 
we further add a supersymmetry breaking term to lead to the discussion in the next section;
\be
S_\mu =\int_{\Sigma_h} dx \, {\cal L}_\mu(x),
\label{SUSY breaking term}
\ee
which is typically assumed to be mass terms. 
For the total action $S+S_\mu$, $K_s(x)$ is given by
\begin{equation}
  K_s(x) = d^\dagger J_s(x) \omega + \bs{s} {\cal L}_\mu(x). 
\end{equation}

Since the integration measure is invariant under 
the supersymmetry transformation $Q$ and $\tilde{Q}$, 
the corresponding WT identities are given by 
\begin{align}
\left\langle
{\cal O}(x_0)  d^\dag J_Q(x)\omega
\right\rangle
&=-\bigl\langle
{\cal O}(x_0)  Q{\cal L}_\mu(x)
\bigr\rangle
+\delta^2(x-x_0)
\bigl\langle
Q{\cal O}(x_0)
\bigr\rangle\,, \\
\left\langle
{\cal O}(x_0)  d^\dag J_{\Qt}(x)\omega
\right\rangle
&=-\bigl\langle
{\cal O}(x_0)  \Qt{\cal L}_\mu(x)
\bigr\rangle
+\delta^2(x-x_0)
\bigl\langle
\Qt{\cal O}(x_0)
\bigr\rangle\,,
\end{align}
respectively.

For the $U(1)_A$ symmetry, we have to be more careful since the integration measure is {\em not} invariant under the $U(1)_A$ transformation.
Using the so-called Fujikawa's method, we see
\begin{equation}
  dX' = dX \exp\left( i\frac{{\rm dim}G}{4\pi}\int_{\Sigma_h} \theta_A(x) R(x)\omega \right) \,,
\end{equation}
where $R(x)$ is the scalar curvature of $\Sigma_h$. 
This is the case where ${\cal A}_s \ne 0$ and is nothing but the $U(1)$ anomaly. 
Thus we obtain the WT identity for the $U(1)_A$ symmetries
\begin{multline}
\left\langle
  {\cal O}(x_0)  \left(d^\dag J_{A}(x) + \frac{{\rm dim}G}{4\pi} R(x) \right)\omega
\right\rangle\\
=-\bigl\langle
{\cal O}(x_0)  \bs{s}_A {\cal L}_\mu(x)
\bigr\rangle
+\delta^2(x-x_0)
\bigl\langle
\bs{s}_A{\cal O}(x_0)
\bigr\rangle\,,
\label{eq:U(1)_A WT}
\end{multline}
where $\bs{s}_A$ is the generators of $U(1)_A$ transformation.
In particular, by integrating \eqref{eq:U(1)_A WT} over $\Sigma_h$ with ${\cal O}=1$, we obtain
\begin{equation}
  \int_{\Sigma_h} \left\langle d^\dagger J_A \omega \right\rangle 
  = -\frac{{\rm dim}G}{4\pi} \int_{\Sigma_h} R\omega = - {\rm dim}G\, \chi_h\,,
  \label{eq:anomalous WT continuum}
\end{equation}
where $\chi_h\equiv 2-2h$ is the Euler characteristics of $\Sigma_h$. 
Therefore we see that no anomaly appears on the torus $T^2$ $(h=1)$. 

On the other hand, 
since the integration measure is invariant under the $U(1)_V$ transformation, 
the $U(1)_V$ symmetry is not anomalous and the corresponding WT identity symmetry is given by
\be
\left\langle
{\cal O}(x_0)  d^\dag J_{V}(x)\omega
\right\rangle
=-\bigl\langle
{\cal O}(x_0)  \bs{s}_V {\cal L}_\mu(x)
\bigr\rangle
+\delta^2(x-x_0)
\bigl\langle
\bs{s}_V{\cal O}(x_0)
\bigr\rangle\,.
\label{eq:U(1)_V WT}
\ee
where $\bs{s}_V$ is the generator of the $U(1)_V$ transformation.

\section{Graph Theory for Discretized Supersymmetric Gauge Theory}
\label{Graph Theory for SUSY YM}

In this section, we consider a discretization of the Riemann surface in order to regularize the supersymmetric gauge theory discussed in the previous section. 
Although the way of the discretization is identical to the model given in \cite{Matsuura:2014kha},
we will reconstruct it from the perspective of graph theory. 

\subsection{Discretized Riemann surface as a graph}
\label{sec:def graph}

To regularize the supersymmetric gauge theory considered in the previous section, 
we divide the Riemann surface into polygons, that is, 
the continuous Riemann surface is approximated by an object consisting of polyhedra glued together without gaps.
We assume that polyhedrons are connected by edges, and that there are no vertices in the middle of the edges.  
We call each polyhedron that constitutes a polygon a face.
As a result, a discretized Riemann surface is labeled by a set of vertices $V=\{v_1,\cdots,v_{n_V}\}$, a set of edges $E=\{e_1,\cdots,e_{n_E}\}$, and a set of faces $F=\{f_1,\cdots,f_{n_F}\}$,
where $n_V$, $n_F$ and $n_F$ are the number of the vertices, edges and faces,
respectively.
For simplicity, we will consider only Riemann surfaces without boundaries in this paper.

\begin{figure}[t]
 \begin{minipage}{0.45\linewidth}
    \begin{center}
      \includegraphics[scale=0.5]{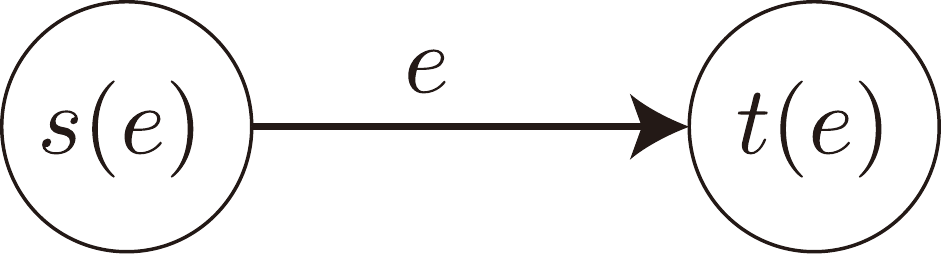}
    \end{center}
      \phantom{blah}
    \caption{A piece of the directed graph.
$s(e)$ and $t(e)$ stand for ``source'' and ``target'' vertices
for a given edge $e$, respectively.}
    \label{a piece of graph}
  \end{minipage}
  \hspace{0.05\linewidth}
  \begin{minipage}{0.45\linewidth}
    \begin{center}
      \includegraphics[scale=0.5]{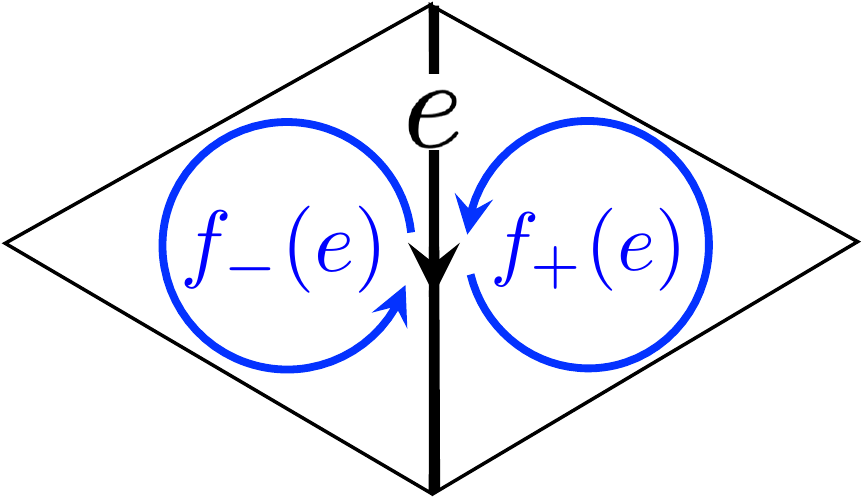}
    \end{center}
    \caption{Another piece of the directed graph. 
$f_+(e)$ and $f_-(e)$ stand for the faces that contain the common edge $e$ 
in the same and the opposite direction for a given edge $e$, respectively.}
    \label{a piece of graph2} 
  \end{minipage}
\end{figure}

The number of faces that a vertex $v$ shares is equal to the number of edges one of whose ends is $v$, 
which we call the degree of the vertex $v$ and denote by ${\rm deg}(v)$.
Similarly, the number of vertices that a face $f$ shares is equal to the number of edges that consist of $f$,
which we call the degree of the face $f$ and denote by ${\rm deg}(f)$.
We can assign a direction to every edge. We thus express the edge $e$ starting from vertex $v_s$ and end to vertex $v_t$ as $e=\{v_s,v_t\}$.
We call $v_s$ the source of $e$ and $v_t$ the target of $e$, and also write $v_s=s(e)$ and $v_t=t(e)$ for given $e$, respectively
(see Fig.~\ref{a piece of graph}).

Since the Riemann surfaces are orientable by definition, we can also consider orientations for the surface $f$. 
Here we adopt the right-handed system and define the direction of the face as the counter-clockwise rotation when we see the Riemann surface from the outside.
Then the surface $f$ can be expressed as $f=\{e_{i_1},\cdots,e_{i_{\rm deg}(f)}\}$ as a list of its constituent edges along the direction of $f$. 
{Since we do not assume the boundary, every} edge is shared by two faces, and these two faces contain a common edge in opposite directions from the way of the construction. 
We then write $f_+(e)$ for the face that contains the edge $e$ in the same direction and $f_-(e)$ for the face that contains it in the opposite direction (See Fig.~\ref{a piece of graph2}).

The observation is that the discretized Riemann surface constructed in this way can be naturally interpreted as a pair of a graph $\Gamma$ and its dual graph $\check\Gamma$. 
A directed graph is defined as a triple $(V,E,\varphi)$ where $V$ is a set of vertices, $E$ is a set of edges, and $\varphi$ is a map $V \times V \to E$.
By considering the relation between $V$ and $E$ constructed above as a map $\varphi:V\times V \to E$, we can naturally regard the triple $(V,E,\varphi)$ as a directed graph $\Gamma$.
Note that, in general graphs, it is possible to draw two or more edges between two definite vertices, or to draw an edge that returns to the same vertex, but the graph we are considering here has at most one edge between two definite vertices, and does not allow edges connecting the same vertex.

On the other hand, the relationship between the faces and edges constructed above can be regarded as a map $\check{\varphi}:F\times F\to E$, thus $(F,E,\check{\varphi})$ can also be regarded as a graph, which is the dual graph $\check\Gamma$ of $\Gamma$.
Therefore, the polygon partition of the Riemann surface considered above can be regarded as a pair $(\Gamma,\check\Gamma)$ of a graph and its dual graph.

\subsection{Matrices describing a graph}
\label{sec:matrices}

We introduce some useful matrices which describe the structures, 
and examine their properties.
These matrices are not only used to define the graph structures, 
but also make it possible to use linear algebra to treat the graph.
In the following, we denote the $n_V$-, $n_E$- and $n_F$-dimensional 
vector spaces on $V$, $E$, and $F$ as ${\cal V}_V$, ${\cal V}_E$ and ${\cal V_F}$,
respectively.
We consider only directed simple graphs in the following. 


\subsubsection*{\underline{\it Incidence matrix }}
The incidence matrix $L$ is a matrix of size $n_E \times n_V$ whose elements are given by 
\be
{L^e}_{v} =
\begin{cases}
+1 &  \text{if $t(e)=v$}\\
-1  & \text{if $s(e)=v$}\\
0  & \text{otherwise}
\end{cases}\,,
\ee
which gives a linear mapping ${\cal V}_V\to{\cal V}_E$ and  can be generated uniquely from the mapping $\varphi:V\times V \to E$ in the triple of the graph.
Note that $L$ is essentially the charge matrix of the quiver theory.
The matrix $L$ acts on a vector ${\bs x} =(x^1,\cdots,x^{n_V})^T \in {\cal V}_V$ as 
\begin{equation}
  L{\bs x} = (x^{t(e_1)}-x^{s(e_1)},\ldots,x^{t(e_{n_E})}-x^{s(e_{n_E})})^T\in {\cal V}_E \,,
\end{equation}
which is a generalization of the forward difference in terms of the lattice gauge theory.
By regarding ${\cal V}_V$ and ${\cal V}_E$ as analogs of the spaces of 0-forms and 1-forms,
$L$ and $L^T$ can be seen as analogs of the exterior derivative $d$ and its adjoint $d^\dagger$ 
in the differential geometry. 


We see that the equation $L{\bs x}=0$ has unique solution $x^{v_1}=\cdots=x^{v_{n_V}}=c$,
since this equation is equivalent to $x^{s(e)}=x^{t(e)}$ and all the vertices are assumed to be connected. 
Therefore, for a connected graph, the rank of the incidence matrix $L$ is $n_V-1$
and $\ker L=\{c {\bs 1}_{n_V}| c\in{\mathbb C}\}$ where ${\boldsymbol 1}_{n_V}=(1,\cdots,1)^T$ in general. 
This is the analog of the fact that the unique solution of $df=0$ is $f={\rm const.}$ in the differential geometry. 
In the following, we denote the normalized zero mode of $L$ as 
\begin{equation}
  \bs{v}_0 \equiv \frac{1}{\sqrt{n_V}} {\bs 1}_{n_V}\,.
  \label{eq:v0}
\end{equation}

To specify $\ker L^T$, we consider a closed loop $C$ made of edges with a direction. 
Correspondingly, we define the vector $\bs{w}_C \in {\cal V}_E$ whose elements are given by 
\begin{equation}
  (\bs{w}_C)^e=
  \begin{cases}
    1  &\text{if $C$ includes $e$ in the same direction} \\
    -1  &\text{if $C$ includes $e$ in the opposite direction} \\
    0  &\text{otherwise}
\end{cases}\,, 
\label{eq:yC}
\end{equation}
which we call the loop vector associated with the loop $C$. 
The loop vector $\bs{w}_C$ satisfies $L^T \bs{w}_C = 0$ since, 
for a fixed vertex $v$ in the loop $C$, 
there are two edges $e_1$ and $e_2$ in $C$ which have the end on $v$ 
and the values of the products 
${{L^T}^{v}}_{e_1}(\bs{w}_C)^{e_1}$ 
and
${{L^T}^{v}}_{e_2}(\bs{w}_C)^{e_2}$ 
are always opposite and cancel with each other from the way of the construction. 
Therefore all the loop vectors $\bs{w}_C$ are elements of $\ker L^T$. 
Furthermore, by counting the dimension, we see that $\ker L^T$ is generated by linearly independent loop vectors. 
First of all, we can construct $n_F$ loop vectors $\bs{w}_f$ associated with each face $f\in F$. 
They are not linearly independent but have one dependence $\sum_{f\in F} \bs{w}_f = 0$, 
so we can construct $n_F-1$ linearly independent loop vectors. 
In addition, there are $2h$ independent non-contractible cycles
on the genus $h$ Riemann surface.
Since the loop vectors $\bs{w}_I$ $(I=1,\cdots,2h)$ corresponding to the cycles
cannot be constructed from $\bs{w}_f$, 
the $n_F+2h-1$ vectors $\{\bs{w}_f, \bs{w}_I\}$ are linearly independent.
Here recalling that the rank of $L$ is $n_V-1$,
we find $\dim\ker L^T = n_E - n_V + 1 =n_F + 2h -1$. 
Therefore, we can conclude that $\{\bs{w}_f, \bs{w}_I\}$ could form a basis of $\ker L^T$.

For the later purpose, we also introduce a matrix
(unoriented incidence matrix) of size $n_E\times n_V$, 
\begin{equation}
  K^e_{\ v} \equiv |L^e_{\ v}| = \begin{cases}
    1 & \text{if $t(e)=v$ or $s(e)=v$} \\
    0 & \text{others}
  \end{cases}\,.
  \label{eq:adjacency matrix}
\end{equation}

\subsubsection*{\underline{\it Dual incidence matrix}}
We can also define the incidence matrix $\check{L}$ for the dual graph $\check\Gamma$, 
which is a matrix of size $n_E \times n_F$ whose elements are given by 
\be
{{\check{L}{}}^e}_{f} =
\begin{cases}
+1 &  \text{if the edge $e$ on the face $f$ is the forward direction}\\
-1  & \text{if the edge $e$ on  the face $f$ is the backward direction}\\
0  & \text{otherwise}
\end{cases}. 
\ee
We call this matrix the dual incidence matrix in the following. 
As well as the incidence matrix, if we restrict restrict ourselves to the relationship between $E$ and $F$, $\check{L}$  and $\check{L}^T$ 
correspond to $d^\dagger$ and $d$ in the differential geometry, respectively. 
By repeating the same discussion as $L$,
we see that $\ker \check{L}$ is the one-dimensional vector space
generated by the constant vector; $\ker \check{L}=\{c {\boldsymbol 1}_{n_F}| c\in{\mathbb C}\}$, 
and $\ker\check{L}^T$ is generated by independent dual loop vectors 
whose dimension is $n_V+2h-1$. 
We denote the normalized zero mode of $\check{L}$ by
\begin{equation}
  \bs{u}_0 \equiv \frac{1}{\sqrt{n_F}} {\bs 1}_{n_F}\,,
  \label{eq:f0}
\end{equation}
as well as the case of the incidence matrix.

We also introduce the unoriented dual incidence matrix by 
\begin{equation}
  \check{K}^e_{\ f} \equiv |\check{L}^e_{\ f}| = \begin{cases}
    1 & \text{if the face $f$ includes the edge $e$} \\
    0 & \text{others}
  \end{cases}\,.
  \label{eq:dual adjacency matrix}
\end{equation}

\subsubsection*{\underline{\it Laplacian matrices}}

The Laplacian matrix (Kirchhoff matrix) $\Delta_V$
acting on the vertex is an $n_V\times n_V$ matrix defined by 
\begin{align}
  ({\Delta_V})^v_{\ v'} &= L^T L = 
\begin{cases}
\deg(v) &  \text{if $v=v'$}\\
-1  & \text{if $v\neq v'$ and $v$ is adjacent to $v'$}\\
0  & \text{otherwise}
\end{cases}\,,
\end{align}
which is also known as the Cartan matrix on the Dynkin diagram (graph) for the Lie algebra.
Note that this matrix is called ``Laplacian'' since it acts on $\bs x$ like 
\be
\bs{x}^T \Delta_V \bs{x} = \sum_{e\in E} (x^{t(e)}-x^{s(e)})^2\,,
\ee
which is nothing but a second order difference operator for the ``field'' $x^v$.
This also supports the analogy $L\leftrightarrow d$ and ${L}^T \leftrightarrow d^\dagger$ because the Laplacian acting on a 0-form $f$ is $\Delta f = d^\dagger d f$ in the differential geometry. 

We also define the face Laplacian matrix $\Delta_F$ by
\begin{align}
  ({\Delta_F})^f_{\ f'} &= \check{L}^T \check{L} =
\begin{cases}
\deg(f) &  \text{if $f=f'$}\\
-1  & \text{if $f$ and $f'$ share the same edge}\\
0  & \text{otherwise}
\end{cases}\,,
\end{align}
and the edge Laplacian by
\begin{equation}
  \Delta_E \equiv LL^T + \check{L} \check{L}^T\,.
\end{equation}

\subsubsection*{\underline{\it Cohomology of $L$ and $\check{L}$ and the eigenvalues of the Laplacian matrices}}
\label{sec:cohomology}

The incidence and dual incidence matrix represent maps from ${\cal V}_V$ to ${\cal V}_E$ and from ${\cal V}_F$ to ${\cal V}_E$, respectively. 
Furthermore, the incidence and dual incidence matrices are orthogonal with each other, 
\be
L^T \check{L} = \check{L}^T L =0,
\label{orthogonality}
\ee
which holds for the same reason that the loop vectors belong to the kernel of $L^T$. 
This is an analog of the nilpotency (exactness) of the differential $d$ and $d^\dagger$. 
This means that we can construct the following exact sequences; 
\be
\begin{split}
&0 \rightarrow {\cal V}_V \xrightarrow{L} {\cal V}_E \xrightarrow{\check{L}^T} {\cal V}_F \rightarrow 0,\\
&0 \leftarrow {\cal V}_V \xleftarrow{L^T} {\cal V}_E \xleftarrow{\check{L}} {\cal V}_F \leftarrow 0.
\end{split}
\ee
Because of this nilpotency, we can define the cohomology group 
$H^V=\ker L$, 
$H^E=\ker \check{L}^T/ \im L$ and 
$H^F={\cal V}_F/ \im \check{L}^T$.

Similar to Hodge's theorem on a compact orientable Riemannian manifold,
the cohomology group is isomorphic to a set of the kernel of the Laplacian (harmonic forms). 
As mentioned in the previous subsection, 
$\dim\ker L =\dim\ker\check{L}=1$ and thus we see
\begin{align}
  \dim\ker \Delta_V &= \dim H^V =1\,, \\
  \dim\ker \Delta_F &=  \dim H^F =1\,.
\end{align}
More explicitly, $\ker \Delta_V$ and $\ker\Delta_F$ are generated by $\bs{v}_0$ and $\bs{u}_0$, respectively. 
Similarly, since $\dim\ker \check{L}^T=n_V+2h-1$ and $\dim{\rm im}L=n_V-1$, we see
\begin{align}
  \dim\ker \Delta_E &= \dim H^E =2h\,.
  \label{eq:dim ker DeltaE}
\end{align}
Note that these results are consistent with the definition of the Euler characteristic
of the graph $\Gamma$,
\be
\chi_h = \dim H^V - \dim H^E + \dim H^F = 2 -2h. 
\ee

In the following argument, the eigenvectors of the Laplacian matrices play important roles. 
Since $\rank L=n_V-1$ and $\rank \check{L}=n_F-1$, 
the rank of $\Delta_V$ and $\Delta_F$ are 
$n_V-1$ and $n_F-1$, respectively. 
Therefore, in addition to the normalized zero modes $\bs{v}_0$ and $\bs{u}_0$ given by 
\eqref{eq:v0} and $\eqref{eq:f0}$, 
$\Delta_V$ and $\Delta_F$ have $n_V-1$ and $n_F-1$ linearly independent eigenvectors with non-zero eigenvalues, respectively. 
We then denote the orthonormal eigenvectors of $\Delta_V$ and $\Delta_F$ 
as $\{\bs{v}_0,\bs{v}_i\}$ $(i=1,\cdots,n_V-1)$ and 
$\{\bs{u}_0,\bs{u}_a\}$ $(a=1,\cdots,n_F-1)$ with 
\be
\begin{split}
&  \Delta_V \bs{v}_0 = 0, \quad \Delta_V \bs{v}_i = \lambda_i \bs{v}_i  \quad (\lambda_i\ne0), \\
&  \Delta_F \bs{u}_0 = 0, \quad \Delta_F \bs{u}_a = \mu_a \bs{u}_a \quad (\mu_a\ne0),
\end{split}
\ee
respectively. 

$L\bs{v}_i$ and $\check{L}\bs{u}_a$ become simultaneously
eigenvectors of $\Delta_E$ with eigenvalues $\lambda_i$ and $\mu_a$, respectively, since
\begin{equation}
\begin{split}
&  \Delta_E (L\bs{v}_i)
=L\Delta_V \bs{v}_i
=\lambda_i (L\bs{v}_i),\\
&  \Delta_E (\check{L}\bs{u}_a)
=\check{L}\Delta_F \bs{u}_a
=\mu_a (\check{L}\bs{u}_a), 
\end{split}
\end{equation}
where we have used the relation \eqref{orthogonality}. 
We then normalize them by
\begin{equation}
 \bs{e}_i \equiv L \bs{v}_i/|L \bs{v}_i|\,, \quad
 \bs{e}_a \equiv \check{L}\bs{u}_a/|\check{L}\bs{u}_a|\,,
\end{equation}
which are orthogonal vectors in ${\cal V}_E$.
To complete the orthonormal basis of ${\cal V}_E$,
we need to add $2h$ independent normalized vectors $\{\bs{e}_0^I\}$ $(I=1,\cdots,2h)$,
which belong to $\ker \Delta_E$,
since $\dim{\cal V}_E=n_E = (n_V-1)+(n_F-1) + 2h$
and $\dim \ker \Delta_E=2h$.
Therefore the orthonormal eigenvectors of $\Delta_E$ are spanned by
$\{\bs{e}_0^I,\bs{e}_i,\bs{e}_a\}$, which satisfy
\begin{align}
  \Delta_E \bs{e}_0^I = 0, \quad 
  \Delta_E \bs{e}_i = \lambda_i \bs{e_i},  \quad 
  \Delta_E \bs{e}_a = \mu_a \bs{e_a}\,.
\end{align}
From the argument below Eq.~\eqref{eq:yC}, 
we find that the zero eigenvectors $\bs{e}_0^I$ correspond 
to the independent non-contractible cycles on the graph $\Gamma$. 
Note also that the non-zero eigenvalues of $\Delta_E$ are common with 
those of $\Delta_V$ and $\Delta_F$, namely
\be
\Spec'\Delta_V \oplus \Spec'\Delta_F = \Spec'\Delta_E,
\ee
where $\Spec'$ stands for a set of the non-zero eigenvalues (spectrum)
of the Laplacian.

\section{Abelian Gauge Theory on the Graph}
\label{Abelian Theory}

In this section, we formulate a supersymmetric Abelian gauge theory on the discretized Riemann surface by using graph theory.
We will see that restricting the theory to Abelian makes it easier to see the zero mode structure of the theory, but will also give us important insight into the relationship between anomalies and zero modes.

\subsection{Definition of the model}

In the following, we will define several vectors on the vertices, edges, and faces, 
which are regarded as fields. 
To define a covariant theory using these fields, we have to consider a structure of metric. 
To this end, we introduce contravariant and covariant vectors, which are expressed 
as vectors with upper and lower indices, respectively.

We consider 
a $n_V$-dimensional bosonic vector
$\bs{\phi}=(\phi^1,\phi^2,\ldots,\phi^{n_V})^T$ and their complex conjugate
$\bar{\bs{\phi}}=(\bar{\phi}^1,\bar{\phi}^2,\ldots,\bar{\phi}^{n_V})^T$, 
whose elements $\phi^v$ and $\bar\phi^v$ are regarded as 
complex bosonic variables (field) living on the vertex $v\in V$.
Note that the position of the indices is important: 
the elements of $\bs{\phi}$ and $\bar{\bs\phi}$ have upper indices 
which indicates that they are contravariant vectors. 
If we take the transpose, they become covariant vectors, who have lower indices 
as 
$\bs{\phi}^T=(\phi_1,\phi_2,\ldots,\phi_{n_V})$ and
$\bar{\bs{\phi}}^T=(\bar{\phi}_1,\bar{\phi}_2,\ldots,\bar{\phi}_{n_V})$.

We also consider a $n_E$-dimensional bosonic vector $\bs{U}=(U^1,U^2,\ldots,U^{n_E})^T$, which will be gauge fields, 
whose elements are assumed to take the values in $U(1)$. 
In this case, we can write $\bs{U}$ by $\exp\left\{i\bs{A}\right\}$,
where $\bs{A}=(A^1,A^2,\ldots,A^{n_E})^T$ whose elements are real variables. 
Furthermore, we introduce an $n_F$-dimensional bosonic vector
$\bs{Y}=(Y^1,Y^2,\dots,Y^{n_F})^T$ whose elements are assumed to be real.   
In addition to the bosonic fields, we also consider the fermionic fields 
$\bs{\eta}=(\eta^1,\eta^2,$
$\ldots,\eta^{n_V})^T\in {\cal V}_V$, 
$\bs{\lambda}=(\lambda^1,\lambda^2,\ldots,\lambda^{n_E})^T\in {\cal V}_E$, 
and $\bs{\chi}=(\chi^1,\chi^2,\ldots,\chi^{n_F})^T\in {\cal V}_F$.

We define the supersymmetry transformations of these fields by
\be
\begin{array}{lcl}
Q \phi^v = 0, && \\
Q\bar{\phi}^v = 2\eta^v, && Q \eta^v=0,\\
QA^e=\lambda^e, && Q \lambda^e = -{L^e}_v \phi^v,\\
QY^f = 0, && Q \chi^f = Y^f.
\end{array}
\label{eq:QSUSY}
\ee
Using this symmetry, we write the action of the model in the $Q$-exact form as 
\be
S = - \frac{1}{2g^2}Q \left\{
\bar{\phi}_v {{L^T}^v}_e\lambda^e
+ \chi_f (Y^f-2\mu(P^f))\right\},
\label{graph Q-exact action}
\ee
where $P^f$ is the plaquette variables 
associated with the face $f$ defined by
\be
P^f \equiv \prod_{e \in f} (U^e)^{{{\check{L}^T}{}^f}_{e}}
=\exp\left\{i{{{\check{L}^T}{}}^f}_e A^e\right\}\,,
\ee
and 
$\mu(P^f)$ is a function called the moment map, which becomes
the field strength $F$ in the continuum limit.

There are several candidates on the moment map. 
The moment map used in the original Sugino model is
\begin{equation}
  \mu(P) = \frac{1}{2i}\frac{P - P^\dagger}{1-\frac{1}{\epsilon^2}||1-P||^2}\,, 
\end{equation}
where $||\cdot||$ is a norm of a matrix defined by $||A||\equiv\sqrt{{\rm tr}(AA^\dagger)}$, 
which is now simply $||A||=|A|^2$ in the Abelian theory, 
and $\epsilon$ is a positive constant parameter chosen in the range $0<\epsilon<2$.  
If we do not introduce the denominator,
the vacuum condition $\mu(P)=0$ has two solutions $P=\pm 1$ which contain an unphysical vacuum at $P=-1$. 
The denominator is necessary to avoid it. See \cite{Sugino:2004qd} for more detail. 

Another candidate is the tangent type function, 
\begin{equation}
  \mu(P) = \frac{1}{2i}\frac{P - P^\dagger}{P+P^\dagger}\,, 
\end{equation}
proposed in \cite{Matsuura:2014pua}. 
$P=-1$ is on the pole of this function and thus $\mu(P)=0$ has a unique solution 
at the physical vacuum.

In addition to them, in Abelian theory, we can use logarithmic type moment map, 
\begin{equation}
  \mu(P)=-i \log(P)\,.
\end{equation}
Although it suffered from the mathematical difficulty of defining the log of a matrix in non-Abelian theory, 
there is no ambiguity in Abelian theory by choosing the branch of the logarithmic function as $-\pi < \arg z \le \pi$ for $z\in{\mathbb C}$.


After eliminating the auxiliary field $Y^f$, the bosonic part of the action becomes
\be
\begin{split}
S_B &= \frac{1}{2g^2} \left\{
\bar{\phi}_v {{L^T}^v}_e {L^e}_{v'} \phi^{v'}
+ \mu(P^f)^2
\right\}\\
&=\frac{1}{2g^2} \left\{
\bar{\bs{\phi}}^T\Delta_V \bs{\phi} + |\bs{\mu}|^2
\right\},
\end{split}
\label{Abelian bosonic action}
\ee
where we have used the definition of the graph vertex Laplacian $\Delta_V = L^TL$
and defined ${\bs \mu}\equiv (\mu(P^{f_1}),\cdots,\mu(P^{f_{n_F}}))^T$.
This action describes the free complex scalar field decoupling from
Maxwell theory in the continuum limit.

The fermionic part of the action becomes
\be
\begin{split}
S_F 
&= -\frac{1}{g^2}\left\{
\eta_v{{L^T}^v}_e \lambda^e
+\chi_f\frac{\delta \mu^f}{\delta A^e}\lambda^e
\right\}\\
&= -\frac{1}{g^2}\left\{
\bs{\eta}^T L^T \bs{\lambda} +\bs{\chi}^T \frac{\delta \bs{\mu}}{\delta \bs{A}}\bs{\lambda}
\right\},
\end{split}
\label{graph fermionic action}
\ee
where $\frac{\delta \bs{\mu}}{\delta \bs{A}}$ is an $n_F \times n_E$ matrix
and is proportional to a transpose of the dual incidence matrix $\check{L}^T$.
Indeed, for each component, we find
\be
\begin{split}
\frac{\delta \mu(P^f)}{\delta A^e}
 &=i\mu'(P^f) P^f {\check{L}^T}{{}^f}_{e},
\end{split}
\ee
where $\mu'$ stands for a derivative of the function $\mu$
w.r.t. the argument.

This model is nothing but the Abelian version of the generalized Sugino model 
constructed in \cite{Matsuura:2014kha}.
Repeating the discussion in \cite{Matsuura:2014kha}, we see that this model is a discretization 
of the ${\cal N}{=}(2,2)$ supersymmetric Abelian gauge theory discussed in Sec.2. 
Because of the discretization, the $\tilde{Q}$-symmetry and the $U(1)_V$ symmetries 
are explicitly broken 
while the $Q$-symmetry and the $U(1)_A$ symmetry are still preserved after discretization at least classically.

\subsection{Fermion zero modes}

Here we will examine the structure of the fermions in this theory. 
We first redefine $\chi^f$ by 
\begin{equation}
  {\rho}^f \equiv i\mu'(P^f)P^f \chi^f,
\end{equation}
where we assume that the prefactor $i\mu'(P^f)P^f$ is non-vanishing and non-singular everywhere\footnote{
The argument in this subsection always holds for $\chi^f$ itself
in the saddle point approximation,
where $\rho^f \sim i \chi^f$ with $P^f\sim 1$.
In the localization method, the zero modes of $\chi^f$ play the same role as
the zero modes of $\rho^f$.
}
.
Then let us combine all fermionic variables into a single $(n_V+n_F+n_E)$-dimensional vector together
such that
\be
\begin{split}
  {{\bs \Psi}} &= 
  \begin{pmatrix}
    {{\bs \eta}}\\
    {\bs \rho}\\
  {{\bs\lambda}}
\end{pmatrix}\,. 
\end{split}
\ee
Then the fermionic part of the action (\ref{graph fermionic action})
can be written as 
\be
\begin{split}
S_F &= \frac{1}{g^2}
\bs{\Psi}^T i\slashed{D} \bs{\Psi},
\end{split}
\ee
where $\slashed{D}$ is a matrix of size $(n_V+n_F+n_E)$, 
\be
\slashed{D} = \begin{pmatrix}
0 & iD^T\\
-iD & 0
\end{pmatrix}
\label{eq:graph Dirac operator}
\ee
with
\be
D = \begin{pmatrix}
L & \check{L}
\end{pmatrix}\,.
\ee

Let us examine the zero modes of the matrix $\slashed{D}$, 
which can be obtained from the zero modes of $D$ and $D^T$. 

$D$ is a linear mapping from ${\cal V}_V\oplus{\cal V}_F$ to ${\cal V}_E$ 
which transforms $(\bs\eta,\bs{\rho})^T\in{\cal V}_V\oplus{\cal V}_F$ as 
\begin{equation}
  D \begin{pmatrix} \bs\eta \\ {\bs\rho} \end{pmatrix} = 
 L{\bs\eta}+\check{L}\bs{\rho}\,. 
\end{equation}
Recalling $\dim \ker L=\dim\ker\check{L}=1$,
we can immediately see that $D$ has one zero mode in $\bs\eta$ and $\bs{\rho}$ each%
\footnote{Note that the equation $L{\bs\eta}+\check{L}\bs{\rho}=0$
leads $L\bs\eta=\check{L}\bs{\rho}=0$ because of the orthogonality 
\eqref{orthogonality}.}
proportional to $\bs{v}_0$ \eqref{eq:v0} and $\bs{u}_0$ \eqref{eq:f0}, respectively. 
Therefore we find $\rank D=n_V+n_F-2 = n_E-2h$
where we have used the definition $\chi_h = n_V - n_E + n_F = 2-2h$.

On the other hand, $D^T$ is a linear mapping 
from ${\cal V}_E$ to ${\cal V}_V\oplus{\cal V}_F$ 
which transforms $\bs\lambda\in{\cal V}_E$ as 
\begin{equation}
  D^T \bs\lambda = 
  \begin{pmatrix} L^T {\bs\lambda} \\ \check{L}^T \bs\lambda \end{pmatrix}\,. 
\end{equation}
Then the zero modes of $D^T$ are the common zero modes of $L^T$ and $\check{L}^T$, 
which are the same as the zero modes $\{\bs{e}_0^I\}$ of $\Delta_E$. 
This is a direct result of the isomorphism $\ker {L}^T/\im \check{L} \simeq \ker\Delta_E$, that is,
$\ker L^T$ is spanned by the $n_F+2h-1$ linearly independent loop vectors 
$\{\bs{e}_0^I,\bs{w}_a\}$ where $\bs{w}_a$ $(a=1,\cdots,n_F-1)$ are 
linearly independent loop vectors corresponding to the faces, 
and 
$\im \check{L}$ is spanned by $\{\bs{w}_a\}$ by construction. 
Therefore $\ker\Delta_E$ is spanned by $\{\bs{e}_0^I\}$.

In summary, $\slashed{D}$ has one zero mode $\eta_{0}$ in ${\bs\eta}$, 
one zero mode $\rho_{0}$ in $\bs{\rho}$ 
and $2h$ zero modes $\lambda_{0}^I$ $(I=1,\cdots,2h)$ in ${\bs\lambda}$, 
which can be explicitly written as 
\begin{align}
  \eta_{0} =  \bs{v}_0\cdot \bs{\eta}, \quad 
  \rho_{0} = \bs{u}_0\cdot\bs{\rho}, \quad
  \lambda_{0}^I = \bs{e}_0^I\cdot \bs{\lambda}\,.
\end{align}

Note that the vertex zero mode $\eta_{0}$ and the edge zero modes $\lambda_{0}^I$ are $Q$-invariant while the face zero mode $\rho_{0}$ is not. 
The $Q$-invariance of $\eta_{0}$ is trivial from \eqref{eq:QSUSY}, 
and $\lambda_{0}^I$ is $Q$-invariant since the loop vector $\bs{e}_0^I$ belongs to $\ker L^T$.
However, $Q$ transforms $\chi_{0}'$ as 
\begin{align}
  Q\rho_{0} &= \frac{1}{\sqrt{n_F}}\sum_{f\in F} Q( i\mu'(P^f) P^f\chi^f ) \nn \\
              &= \frac{1}{\sqrt{n_F}}\sum_{f\in F}\left\{-\left( \mu''(P^f)(P^f)^2+\mu'(P^f)P^f\right)
              \lambda^e \check{L}^e_{\ f} \chi^f + i\mu'(P^f)P^f Y^f\right\}\,,
              \label{eq:Qchi0}
\end{align}
which does not vanish in general.

\subsection{Heat kernel and dimensionality}

We can examine the analytic behavior of the fermion spectrum by using the hear kernel.
In the continuous theory, the heat kernel is defined by the following heat equation;
\be
\left(\frac{\del}{\del t}+\slashed{\cal D}^2\right)h(x,y;t)=0,
\ee
which is formally solved as
\be
h(x,y;t) = e^{-t \slashed{\cal D}^2}\,. 
\ee
In particular, the regularized $U(1)_A$ current is evaluated by the heat kernel;
\be
\int_{\Sigma_h}\left\langle
d^\dag J_A \omega 
\right\rangle_{\text{reg}}
=-\Tr \gamma_A e^{-t \slashed{\cal D}^2}
=-\ind \slashed{\cal D}\,.
\label{cont index}
\ee

We can extend the definition of the heat kernel to the Dirac operator on the graph.
The heat kernel on the graph is simply defined by
\be
{h(t)^i}_j \equiv {\left(e^{-t \slashed{D}^2}\right)^i}_j,
\ee
using the matrix $\slashed{D}$ defined by \eqref{eq:graph Dirac operator}. 
Correspondingly, we define the quantity, 
\be
\begin{split}
  I(t) &\equiv \Tr_{V\oplus F\oplus E} \gamma_A e^{-t \slashed{D}^2} \\
&=\Tr_V e^{-t \Delta_V} + \Tr_F e^{-t \Delta_F} - \Tr_E e^{-t \Delta_E}\\
&=\sum_{l=1}^{n_V} e^{-t \lambda^V_l}
+ \sum_{m=1}^{n_F} e^{-t \lambda^F_m}
-\sum_{n=1}^{n_E} e^{-t \lambda^E_n},
\end{split}
\label{index I(t)}
\ee
where $\lambda^V_l$, $\lambda^F_m$ and $\lambda^E_n$
are eigenvalues for the Laplacians $\Delta_V$, $\Delta_F$ and $\Delta_E$,
respectively.

In the large $t$ limit, the contribution of the non-zero modes (eigenvalues)
to $I(t)$ disappears and the difference of the number of the zero modes survives as the index.
Moreover, the contributions from the non-zero modes are canceled with each other
even when the value of $t$ is finite 
because there is a one-to-one correspondence between the non-zero modes 
of $\{\Delta_V,\Delta_F\}$ and $\Delta_E$ as shown in Sec.~\ref{sec:cohomology}. 
Therefore $I(t)$ is independent of $t$, 
that is, 
$I(t)$ gives the index of $\slashed{D}$ which is equal to the Euler characteristic of the graph,
as well as in the continuous theory.

It is instructive to give some concrete examples of the trace of heat kernels.
The simplest example of graph with genus zero is the tetrahedron ($(n_V,n_F,n_E)=(4,4,6)$). 
The eigenvalues of the Laplacians are given by
\be
\begin{split}
\Spec \Delta_V &= \{4,4,4,0\},\\
\Spec \Delta_F &= \{4,4,4,0\},\\
\Spec \Delta_E &= \{4,4,4,4,4,4\}.
\end{split}
\ee
Then we get
\be
\Tr \gamma_A e^{-t \slashed{D}^2}
=(3e^{-4t}+1)+ (3e^{-4t}+1) - 6e^{-4t} = 2, 
\ee
where the non-zero modes are canceled with each other order by order.

Similarly, for the hexahedron ($(n_V,n_F,n_E)=(8,6,12)$), we obtain
\be
\begin{split}
\Spec \Delta_V &= \{6,4,4,4,2,2,2,0\},\\
\Spec \Delta_F &= \{6,6,4,4,4,0\},\\
\Spec \Delta_E &= \{6,6,6,4,4,4,4,4,4,2,2,2\}\,,
\end{split}
\ee
then we obtain
\begin{multline}
\Tr \gamma_A e^{-t \slashed{D}^2}
=(e^{-6t}+3e^{-4t}+3e^{-2t}+1)\\
+ (2e^{-6t}+3e^{-4t}+1)
 - (3e^{-6t}+6e^{-4t}+3e^{-2t})
 = 2.
\end{multline}

For genus 1, the spectrum of the Laplacians of the $3\times 3$ torus is 
\be
\begin{split}
\Spec \Delta_V &= \{6,6,6,6,3,3,3,3,0\}\,,\\
\Spec \Delta_F &= \{6,6,6,6,3,3,3,3,0\}\,,\\
\Spec \Delta_E &= \{6,6,6,6,6,6,6,6,3,3,3,3,3,3,3,3,0,0\}\,, 
\end{split}
\ee
and the index becomes  
\be
\begin{split}
\Tr \gamma_A e^{-t \slashed{D}^2}
&=(4e^{-6t}+4e^{-3t}+1)
+ (4e^{-6t}+4e^{-3t}+1)
 - (8e^{-6t}+8e^{-3t}+2)\\
 & = 0.
\end{split}
\ee
as expected. 

Furthermore, 
the behavior of the heat kernel for each Laplacian (not square of the Dirac operator)
also represents the dimensionality of the graph structure.
If we take the continuum limit of the graph discretization (lattice), 
we expect that the space-time goes to the smooth Riemann surface.
The heat kernel on the Riemann surface,
which satisfies the heat equation
\be
\left(\frac{\del}{\del t}+\Delta_x\right)h(x,y;t)=0\,,
\ee
with the Laplacian $\Delta_x$,
 behaves as
\be
h(x,y;t) = \frac{1}{4\pi t}e^{-|x-y|/2t}+\cdots,
\ee
for small $t$,
while the index (\ref{cont index}) is independent of $t$.
In particular, the trace of the heat kernel behaves as
\be
\tilde{h}(t) \equiv  \int_{\Sigma_h} dx \, h(x,x;t) = \frac{{\rm Vol}(\Sigma_h)}{4\pi t}+ \cdots\,,
\ee
and, in the large $t$ limit, the trace of the heat kernel tends to the number of the
zero modes
\be
\lim_{t\to \infty} \tilde{h}(t) =\lim_{t\to \infty}  \sum_n e^{-t \lambda_n}
=\dim \ker \Delta_V\,.
\ee

On the general $D$-dimensional space-time, the trace of the heat kernel is proportional
to $1/t^{D/2}$ for the small $t$. So, if we investigate the small $t$ behavior of the
heat kernel for the graph Laplacian,
we can confirm how the dimensionality of the
graph discretization is close to the  Riemann surface.

\begin{figure}[t]
\begin{center}
\includegraphics[scale=0.9]{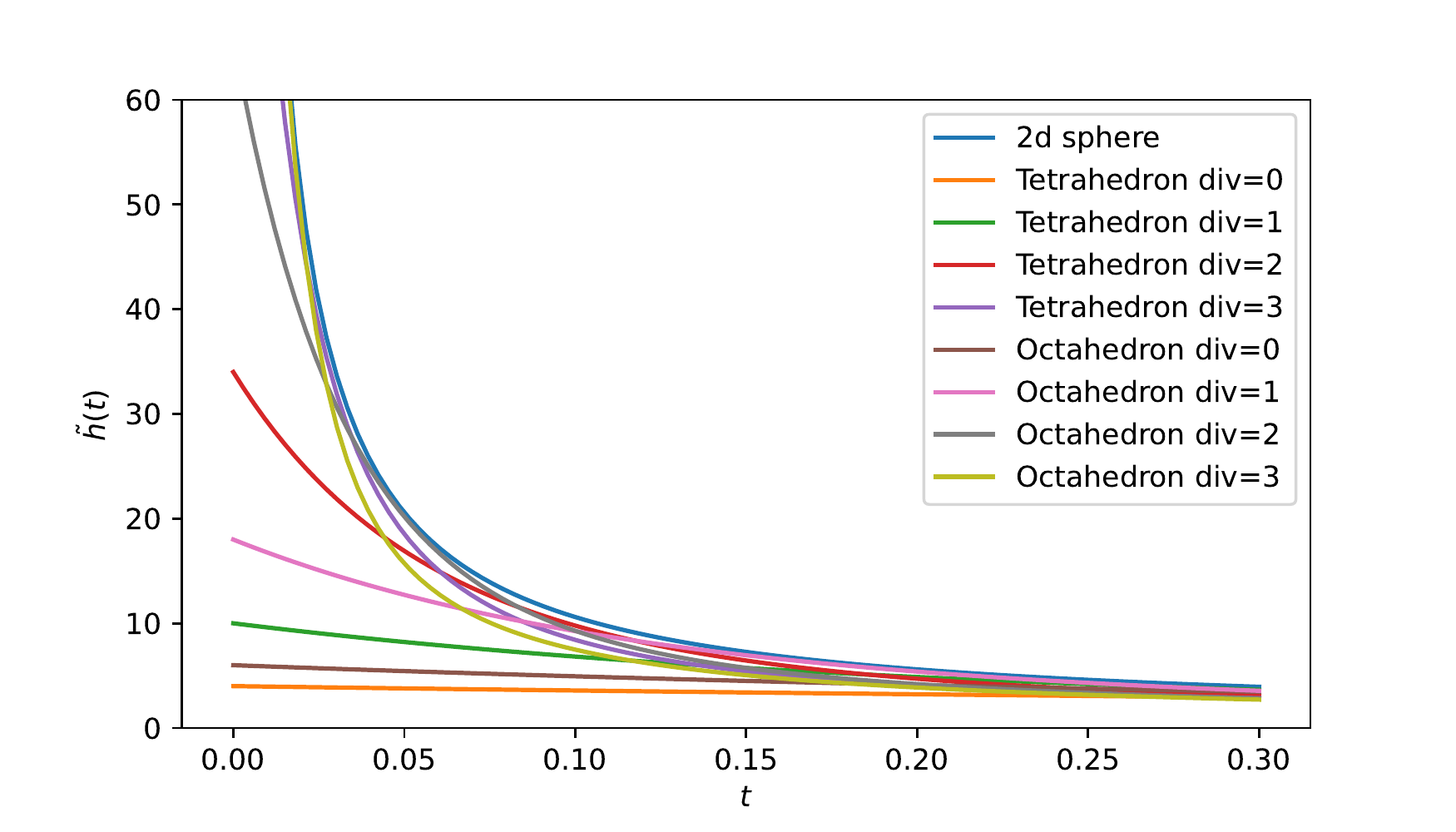}
\end{center}
\caption{Plots of the heat kernel. If the number of vertices
(or faces) of the geodesic polyhedra
is large, the behaviour of the heat kernel becomes very close to that of the 
smooth two-dimensional sphere.
(The ``div'' in the legend stands for the number of times
the triangular faces of the tetrahedron or
octahedron is divided into smaller triangles.)
}
\label{heat kernel}
\end{figure}

We construct the heat kernels of the graph Laplacian for several
geodesic polyhedrons with genus 0 (subdivisions of tetrahedron and octahedron) 
and plot in Fig.~\ref{heat kernel}.
We find that 
the behavior of the heat kernel approached that of the two-dimensional sphere
($1/t$ behavior)
as the number of the vertices increases and the discretization becomes finer.
We also find that the trace of the heat kernel represents the number of the
zero modes in the large $t$ limit.

\subsection{Uplifting the fermion zero modes}

Due to the existence of these zero modes, the discretized theory 
with the action \eqref{graph Q-exact action} is not well-defined 
since the partition function trivially vanishes. 
we expand the fermion fields by the eigenvectors of the Laplacian matrices as 
\begin{align}
  \bs{\eta}&=\eta_0 \bs{v}_0 + \sum_{i=1}^{n_V-1} \eta_i \bs{v}_i, \quad
  \bs{\chi}=\chi_0 \bs{u}_0 + \sum_{a=1}^{n_F-1} \chi_a \bs{u}_a, \nn \\
  \bs{\lambda}&=\sum_{I=1}^{2h} \lambda_0^I \bs{e}_0^I 
  + \sum_{i=1}^{n_V-1} \lambda_i \bs{e}_i
  + \sum_{a=1}^{n_F-1} \lambda_a \bs{e}_a\,,
\end{align}
and write the integration measure for the modes as 
$d{\cal B}d{\cal F}d{\cal F}_0$,
where
\begin{align}
  d{\cal B} &= \left(\prod_{v=1}^{n_V}d\phi_v d\bar\phi_v\right)
  \left(\prod_{e=1}^{n_E} dU_e\right), \\
  d{\cal F} &= \left(\prod_{i=1}^{n_V-1} d\eta_i d\lambda_i \right)
  \left(\prod_{a=1}^{n_F-1} d\chi_a d\lambda_a \right), \\
  d{\cal F}_0 &= 
  \left(\prod_{I=1}^{2h} d\lambda_0^{2h-I+1}\right) d\chi_0 d\eta_0 \,.
  \label{eq:zero mode measure}
\end{align}
Evaluating the vev of a (not necessarily local) operator ${\cal O}[X]$,
which
is a functional of the collective expression of all the fields $X$
and
does not include any fermion zero mode, 
the integration $\int d{\cal B}d{\cal F}d{\cal F}_0\, {\cal O}[X] e^{-S[X]}$ trivially vanishes.

To avoid it, we have to insert all the fermion zero modes in the background like 
\begin{equation}
  {\cal I}_{\cal O} \equiv \int d{\cal B}d{\cal F}d{\cal F}_0
  \left(\eta_{0}\chi_{0}\prod_{I=1}^{2h} \lambda_{0}^I\right) \,
  {\cal{O}}[X]e^{-S[X]}\,. 
  \label{eq:IO}
\end{equation}
The inserted zero modes are integrated
by the measure of $d{\cal F}_0$, then
\begin{equation}
  {\cal I}_{\cal O} =\int d{\cal B} d{\cal F}\, {\cal O}[X]
e^{-S[X]}\,, 
\end{equation}
becomes well-defined since the measure does not include the fermion zero modes anymore.

The straightforward way to achieve it automatically is 
to add mass terms of the fermion zero modes to the action as 
\begin{align}
  S_\mu &= \frac{\mu_0}{g^2}\left( \eta_{0}\chi_{0} \right)
  + \frac{\mu_1}{g^2} \sum_{k=1}^h \lambda_{0}^{2k-1}  \lambda_{0}^{2k} 
        \equiv \frac{1}{2g^2} {\bs\Psi}^T 
        {\cal M}{\bs\Psi}
\,.
\label{eq:mass term}
\end{align}
These terms not only make the Dirac matrix invertible 
but also supply the necessary fermion zero modes in evaluating correlation functions as 
\begin{equation}
  \int d{\cal B}d{\cal F}d{\cal F}_0
  \, {\cal{O}}[X]e^{-S[X]-S_\mu} = 
  \frac{\mu_0 \mu_1^h}{g^{2h+2}}\,
  {\cal I}_{\cal O}\,.
  \label{eq:supply}
\end{equation}
The necessary fermion zero modes are supplied by expanding $e^{-S_\mu}$ by the parameters $\mu_0$ and $\mu_1$ as the term with the coefficient $\mu_0\mu_1^h$, 
and the other terms vanish as lack or excess of the fermion zero modes. 
Note that the situation is the same when the operator ${\cal O}$ includes all or a part of the fermion zero modes, where only the term including all the fermion zero modes survives. 

Here we note that the supplied zero modes from $e^{-S_\mu}$ break the $U(1)_A$ symmetry unless $h=1$ 
reflecting the quantum anomaly discussed soon later. 
We also note that the mass terms (\ref{eq:mass term})
also break the $Q$-symmetry softly 
since $\chi_0$ is not $Q$-invariant as shown in \eqref{eq:Qchi0}.
As we will discuss in the next section, 
it is possible to construct such mass terms that cancel the fermion zero modes 
while preserving the $Q$-symmetry. 
In this sense, although the mass term constructed here is simple, 
it is not the only option. 
In particular, when discussing situations where supersymmetry 
plays an important role, 
we should use the $Q$-invariant mass terms. 
However the choice of mass term is not so important 
to discuss the anomalous $U(1)_A$ current, 
as long as the partition function is well-defined. 
Therefore, in this section,
we will use 
the regularized action, 
\begin{align}
S &= \frac{1}{2g^2} \left\{
\bar{\bs{\phi}}^T\Delta_V \bs{\phi} + |\bs{\mu}|^2
+ {\bs\Psi}^T \left(i\slashed{D} + {\cal M} \right) {\bs\Psi}
\right\}\,,
\end{align}
for a while.

\subsection{Chiral anomaly on the graph}

We next consider the WT identity corresponding to the $U(1)_A$ symmetry. 
To this end, we consider the following local $U(1)_A$ transformation,
\begin{equation}
  \eta^v \to e^{-i\theta_{\!A}^v} \eta^v, \quad
  \lambda^e \to e^{i\theta_{\!A}^e} \lambda^e, \quad
  \chi^f \to e^{-i\theta_{\!A}^f} \chi^f. 
  \label{eq:U1A local}
\end{equation}
This local transformation is not the symmetry of the theory in general,
but it makes the action invariant if the transformation is global
(independent of the positions),
namely
$\theta_{\!A}^v=\theta_{\!A}^e=\theta_{\!A}^f$.
We are now dealing with the graph $\Gamma$ and the dual graph $\check\Gamma$ on an equal footing, 
where one edge is shared by two vertices and two faces. 
To respect this structure, we take the transformation parameters $\theta_{\!A}^e$ of the edge $e$ to be the average of the transformation parameters of the vertices $\theta_{\!A}^v$ and faces $\theta_{\!A}^f$ as 
\begin{equation}
  \theta_{\!A}^e=\frac{1}{4}(\theta_{\!A}^{s(e)}+\theta_{\!A}^{t(e)}+\theta_{\!A}^{f_+(e)}+\theta_{\!A}^{f_-(e)})\,,
  \label{eq:constraint}
\end{equation}
where $s(e)$, $t(e)$ and $f_\pm(e)$
are defined in Sec.~\ref{sec:def graph}.
Then the infinitesimal transformation of the action becomes
\begin{align}
  \delta S 
  &= \sum_{e\in E} \Bigl\{
    (L^e_{\ v} \theta_{\!A}^v) J^{(V)}_e 
    +(\check{L}^e_{\ f}\theta_{\!A}^f) J^{(F)}_e 
    +\left( (K^e_{\ v} \theta_{\!A}^v) - (\check{K}^e_{\ f}\theta_{\!A}^f)\right) G_e
  \Bigr\}\,, 
  \label{eq:delta S}
\end{align}
where $K$ and $\check{K}$ are the matrices 
given by \eqref{eq:adjacency matrix} and \eqref{eq:dual adjacency matrix}, respectively, and 
\begin{align}
  J^{(V)}_e &\equiv 
  \frac{i}{g^2}\Bigl(
  -\bar\phi^{t(e)} \phi^{s(e)} +  \bar\phi^{s(e)}\phi^{t(e)} 
  + \frac{1}{2} \lambda_e \sum_{v\in V} K^e_{\ v} \eta^v
\Bigr)\,, \\
  J^{(F)}_e &\equiv 
  \frac{i}{2g^2} \lambda_e \sum_{f\in F}\check{K}^e_{\ f} \hat\chi^f \,, \\ 
  G_e &\equiv \frac{i}{4g^2}\lambda_e \left( 
    \sum_{v\in V} (L^e_{\ v} \eta^v)
    -\sum_{f\in F} (\check{L}^e_{\ f} \hat\chi^f )
    \right)\,, 
\end{align}
which are defined on each edge $e\in E$ and thus the index $e$ is not contracted. 

Recalling that the mass terms \eqref{eq:mass term} effectively supply all the fermion zero modes to the integration, we have also to evaluate the transformation of the integral measure $d{\cal B}d{\cal F}d{\cal F}_0$ and the inserted fermion zero modes. 
The infinitesimal transformation of the measure becomes
\begin{equation}
  d{\cal B}d{\cal F}d{\cal F}_0 \to 
  \left( 1 +  \sum_v (i\theta_{\!A}^v)(1-\frac{1}{4}{\rm deg}(v))
  +\sum_f (i\theta_{\!A}^f)(1-\frac{1}{4}{\rm deg}(f))\right)
  d{\cal B}d{\cal F}d{\cal F}_0 \,,
\end{equation}
and that of each fermion zero mode is 
\begin{align}
  \eta_0 &\to \eta_0 -i \sum_{v\in V}\theta_{\!A}^v (\bs{v}_0)_v \eta^v, \nn \\
  \chi_0 &\to \chi_0 -i \sum_{f\in F}\theta_{\!A}^f (\bs{u}_0)_f \hat\chi^f,  \\
  \lambda_0^I &\to \lambda_0^I +i \sum_{e\in E}\theta_{\!A}^d (\bs{e}_0^I)_e \lambda^e\,. \nn
\end{align}
So the zero mode integral is evaluated as
\begin{align}
  \int d{\cal F}_0\, \delta\left(\eta_0 \chi_0 \prod_{I=1}^{2h} \lambda_0^I \right)
  = &\sum_{v\in V} (i\theta_{\!A}^v)\left( -(\bs{v}_0)_v^2 + \frac{1}{4}\sum_{I=1}^{2h}\sum_{e\in E} K^{e}_{\ v} (\bs{e}_0^I)_e^2 \right)\nn \\ 
  +& \sum_{f\in F} (i\theta_{\!A}^f)\left( -(\bs{u}_0)_f^2 + \frac{1}{4}\sum_{I=1}^{2h}\sum_{e\in E} \check{K}^{e}_{\ f} (\bs{e}_0^I)_e^2 \right)\,. 
    \label{eq:delta dXF0}
\end{align}

Combining \eqref{eq:delta S} and \eqref{eq:delta dXF0}, we obtain the identities,
\begin{align}
\label{eq:id1}
\biggl\langle 
  \sum_{e\in E}\left( L^e_{\ v}J_e^{(V)} + K^e_{\ v} G_e \right)
\biggr\rangle
&= -\left(1 - \frac{1}{4} {\rm deg}(v)\right)
+\left((\bs{v}_0)_v^2 - \frac{1}{4}\sum_{I=1}^{2h}\sum_{e\in E} K^{e}_{\ v} (\bs{e}_0^I)_e^2\right)  \,, \\
\biggl\langle 
  \sum_{e\in E} \left( \check{L}^{e}_{\ f}J^{(F)}_e - \check{K}^{e}_{\ f}G_e \right)
\biggr\rangle
&= -\left(1 - \frac{1}{4} {\rm deg}(f)\right)
+\left( (\bs{u}_0)_f^2 - \frac{1}{4}\sum_{I=1}^{2h}\sum_{e\in E} \check{K}^{e}_{\ f} (\bs{e}_0^I)_e^2 \right) \,,
\label{eq:id2}
\end{align}
on each $v\in V$ and $f\in F$, respectively. 
These identities are the local WT identities corresponding to \eqref{eq:U(1)_A WT} in the continuous theory.
Note that 
we have two local WT identities for the single $U(1)_A$ symmetry
because we are considering the graph and the dual graph at the same time: 
The transformation parameters $\theta_{\!A}^v$ and $\theta_{\!A}^f$ 
can be assigned independently on the vertices of the graph and the dual graph, respectively. 

Interestingly, 
the quantities $1-{\rm deg}(v)/4$ and $1-{\rm deg}(f)/4$ appearing 
in the first parentheses of the right-hand sides 
correspond to the scalar curvature of the continuum geometry. 
We can justify it from the fact that they can be regarded as the deficit angle, 
which is proportional to the scalar curvature in two-dimensional geometry. 
To see it, 
suppose that all the faces are regular $n$-polygon of the same size. 
In this case ${\rm deg}(f)=n$ and 
the deficit angle around the vertex $v$ is given by 
$\theta_v=2\pi\left(1-\frac{n-2}{2n}{\rm deg}(v)\right)$. 
Then the average of the quantities around a vertex $v$ becomes the deficit angle as announced;
\begin{equation}
  1-\frac{1}{4}{\rm deg}(v) + \sum_{f\in F_v}\frac{1}{{\rm deg}(f)}  \left(1-\frac{1}{4}{\rm deg}(f)\right) = \frac{\theta_v}{2\pi}\,,
  \label{eq:average}
\end{equation}
where $F_v$ denotes the faces touching at the vertex $v$.

The second parentheses of the right-hand sides of 
\eqref{eq:id1} and \eqref{eq:id2}
are the contribution of the inserted fermion zero modes, 
which are necessary so that these identities are consistent. 
We can see it as follows: 
If we take summation over all the vertices of \eqref{eq:id1} and all the faces of \eqref{eq:id2} followed by summing these two identities, 
the terms with $G_e$ trivially cancel and the remaining 
terms $\sum_{v,e} L^e_{\ v} J^{(V)}_e$ and $\sum_{f,e} {\check{L}}^e_{\ f} J^{(F)}_e$
vanish because of the structure of the matrices $L$ and $\check{L}$. 
This corresponds to the fact that the integration over a total derivative vanishes in the continuous theory. 
On the other hand, after the same operation, the first term of the right-hand side gives
\begin{align}
-n_V - n_F  + \frac{1}{4} \sum_v {\rm deg}(v) - \frac{1}{4} \sum_f {\rm deg}(f) 
= -n_V - n_F + n_E 
= -\chi_h, 
\label{eq:rhs1}
\end{align}
which is a reproduction of the anomalous WT identity 
\eqref{eq:anomalous WT continuum} in the continuous theory with $G=U(1)$. 
However, if the second terms of the right-hand sides are absent,
we obtain an inconsistent expression unless $\chi_h=0$ because the left-hand side vanishes. 
The second terms cure the situation
such that
\begin{align}
  \sum_{v\in V} 
\left((\bs{v}_0)_v^2 - \frac{1}{4}\sum_{I=1}^{2h}\sum_{e\in E} K^{e}_{\ v} (\bs{e}_0^I)_e^2\right)
+ \sum_{f\in F}
\left( (\bs{u}_0)_f^2 - \frac{1}{4}\sum_{I=1}^{2h}\sum_{e\in E} \check{K}^{e}_{\ f} (\bs{e}_0^I)_e^2 \right) = 2-2h = \chi_h\,,
\label{eq:rhs2}
\end{align}
which cancels the contribution from the first terms \eqref{eq:rhs1}. 
This is consistent with the fact that the left-hand side vanishes.

\section{Localization in the Abelian Theory}
\label{Localization in the Abelian theory}
\subsection{Saddle point equation}

We now apply the localization method to perform the path integral
of the Abelian theory exactly.

Let $f[X]$ be  a $Q$-closed operator.
Recalling the action is written in the $Q$-exact form
such that $S=Q \Xi$,
we see that the integration
\be
\left\langle f[X]\right\rangle_t \equiv 
\int dX \, f[X]e^{-tS}
\ee
is independent of $t$.
If we differentiate it by $t$, we obtain 
the vev of a $Q$-exact operator,
which vanishes at a supersymmetric vacuum;
\be
\frac{\del}{\del t} \langle f[X] \rangle_t = -\left\langle Q(f[X]\, \Xi)
\right\rangle_t=0\,.
\ee
The path integral is localized at the saddle points in the limit of $t \to \infty$. 
Thus we can evaluate the vev of a $Q$-closed operator, 
which includes the partition function itself, exactly 
by the saddle point approximation of the Abelian theory. 

From the bosonic action (\ref{Abelian bosonic action}),
we can see the saddle points are given by equations
\be
L \bs{\phi} = 0,
\quad
\bs{\mu} = 0.
\ee
On the connected graph, the former equation
always has a solution
\be
\bs{\phi} = \phi_0 \bs{v}_0
\ee
where
$\phi_0 \in \mathbb{C}$ and  $\bs{v}_0$ is the zero mode of $L$ given by \eqref{eq:v0}. 
We then expand the scalar fields around the saddle point as 
\be
\begin{split}
\bs{\phi} &= \phi_0 \bs{v}_0 + \frac{1}{\sqrt{t}}\tilde{\bs{\phi}},\\
\bar{\bs{\phi}} &= \bar{\phi}_0 \bs{v}_0 + \frac{1}{\sqrt{t}}\tilde{\bar{\bs{\phi}}},\\
\end{split}
\label{eq:expansion}
\ee
where we have rescaled the fluctuation by $1/\sqrt{t}$ for the later purpose. 

For the  moment map $\mu(P^f)$,
the latter saddle point equation just means $P^f = 1$.
Since the plaquette variable is given by
\be
P^f = \exp\left\{
i {{L^T}^f}_e A^e
\right\},
\ee
the moment map constraint is solved by
\be
 {{\check{L}^T}{}^f}_{e}\hat{A}^e = 2\pi k^f,
 \label{eq:fixed pt of A}
\ee
where $k^f$ are integer numbers.
As same as the scalar fields, we expand the gauge field around the fixed point as 
\begin{equation}
\bs{A} = \hat{\bs{A}} + \frac{1}{\sqrt{t}}\tilde{\bs{A}}\,. 
\end{equation}
In particular, around the fix points, the plaquette variable approximately behaves as 
\be
\bs{\mu} \sim \frac{1}{\sqrt{t}} \check{L}^T\tilde{\bs{A}},
\ee
up to a linear order of a fluctuation $\tilde{\bs{A}}$.
Then the face part of the bosonic action around the saddle point becomes 
\begin{equation}
  \bs{Y}^T \cdot \left( \bs{Y} - \frac{2}{\sqrt{t}} \check{L}^T \tilde{\bs{A}}\right)\,. 
  \label{eq:Sb face}
\end{equation}
Looking at this expression, we see that the trace mode in the auxiliary field $\bs{Y}$ is decoupled from the gauge field at the saddle point. 
We then separate it from the others as 
\be
\bs{Y}=\frac{1}{\sqrt{t}} \left( Y_0\bs{u}_0+\tilde{\bs{Y}} \right)\,,
\ee
and rewrite \eqref{eq:Sb face} as
\begin{equation}
  \frac{1}{t} \left\{
  Y_0^2 + \tilde{\bs{Y}}^T\cdot \left( \tilde{\bs{Y}} - 2\check{L} \tilde{\bs{A}} \right) \right\}\,.
\end{equation}
Note that we have put the factor $\frac{1}{\sqrt{t}}$ also to $Y_0$ because 
it is not the zero mode in the action but is just a free mode as we have seen above.

After integrating out $\tilde{\bs{Y}}$, 
the 1-loop effective bosonic action becomes
\be
S_B^{\text{1-loop}}\equiv
\lim_{t\to\infty}tS_B 
=  \frac{1}{2g^2} \left\{
-Y_0^2
+\tilde{\bar{\bs{\phi}}}^T\Delta_V \tilde{\bs{\phi}}
+ \tilde{\bs{A}}^T\check{L}\check{L}^T \tilde{\bs{A}}
\right\}\,.
\label{1-loop effective Abelian action for bosons}
\ee
We can ignore the first term in the present localization argument,
since the integral of $Y_0$ is just Gaussian
(by a rotation $Y_0\to iY_0$) and irrelevant
if any operator coupled with $\bs{Y}$ is not inserted.
However, we will see later that the existence of $Y_0$ becomes important 
when inserting suitable operators including the auxiliary field.

For the fermions, we expand the fields around the fermion zero modes (if they exist) as
\begin{align}
  \bs{\eta} = {\eta_0}\bs{v}_0 + \frac{1}{\sqrt{t}}\tilde{\bs{\eta}}, \quad
  \bs{\chi} = {\chi_0}\bs{u}_0 + \frac{1}{\sqrt{t}}\tilde{\bs{\chi}}, \quad
  \bs{\lambda} = \sum_{I=1}^{2h}{\lambda_0^I}\bs{e}_0^I + \frac{1}{\sqrt{t}}\tilde{\bs{\lambda}}\,. 
  \label{eq:expansion-f}
\end{align}
Then the 1-loop effective action for the fermions
reduces to
\be
S_F^{\text{1-loop}}\equiv
\lim_{t\to\infty}tS_F
=
-\frac{1}{g^2}\left\{
\tilde{\bs{\eta}}^T L^T \tilde{\bs{\lambda}}
+\tilde{\bs{\chi}}^T \check{L}^T \tilde{\bs{\lambda}}
\right\}.
\label{1-loop effective Abelian action for fermions} 
\ee
Note here that the fermion zero modes do not appear at all in the effective action
since they are defined as the kernels of $L$ and $\check{L}$.

For the inserted operator $f[X]$, 
by inserting the expansion \eqref{eq:expansion} 
and \eqref{eq:expansion-f}, 
we obtain
\begin{equation}
  \lim_{t\to\infty} f[X] = f[X_0]\,,
\end{equation}
where $X_0$ is the collective expression of the variables of the saddle points. 
Therefore, in evaluating the integration, 
we have only to consider the inserted operator only 
in integrating over $X_0$. 


\subsection{Gauge fixing and 1-loop determinant}

The gauge transformation of the edge variable $A^e$ is given by
\be
\bs{A}' = \bs{A} - L \bs{\xi},
\ee
where $\bs{\xi}\in {\cal V}_V$ is a gauge transformation parameter.
We can see immediately (the exponent of) the plaquette variable
\be
\bs{P} = \exp\left[i \check{L}^T \bs{A}\right]
\ee
 is invariant under this transformation because of the orthogonality
 of the incidence matrices (\ref{orthogonality}).
This gauge invariance still keeps in the 1-loop
effective action (\ref{1-loop effective Abelian action for bosons})
and 
(\ref{1-loop effective Abelian action for fermions}).

To proceed with the quantization (path integral) of the
1-loop effective theory, we introduce a gauge fixing condition
for the fluctuations
\be
L^T \tilde{\bs{A}} = 0,
\ee
which is an analogy with the Lorentz gauge in the continuous theory.

Introducing the Faddeev-Popov (FP) ghosts $(\bs{c},\bar{\bs{c}})$ and the Nakanishi-Lautrup (NL) field $\bs{B}$ defined on $V$ 
obey the BRST transformations
\be
\delta_B \bs{c} = 0,\quad
\delta_B \bar{\bs{c}} = 2\bs{B},\quad
\delta_B \bs{B}=0,
\ee
the gauge fixing and FP term is written by a BRST exact form
\be
\begin{split}
S_{\rm{GF+FP}} &= -\frac{1}{4g^2}\delta_B \left\{
\bar{\bs{c}}^T \cdot \left(\bs{B}-2L^T\tilde{\bs{A}}\right)
\right\}\\
&=\frac{1}{2g^2}\left\{
\bar{\bs{c}}^T\Delta_V \bs{c}
-\bs{B}^T\cdot\left(\bs{B}-2L^T\tilde{\bs{A}}\right)
\right\},
\end{split}
\ee
where $\Delta_V=L^TL$ and we have used the BRST transformation of the gauge field
\be
\delta_B \tilde{\bs{A}} = -L\bs{c}.
\ee

Combining the original action and gauge fixing terms, we obtain the total action
\be
\begin{split}
S' &= S^{\text{1-loop}}_B + S^{\text{1-loop}}_F + S_{\rm{GF+FP}}\\
&= \frac{1}{2g^2}\Big\{
\tilde{\bar{\bs{\phi}}}^T\Delta_V \tilde{\bs{\phi}}
+ \tilde{\bs{A}}^T\check{L}\check{L}^T \tilde{\bs{A}}
-2\tilde{\bs{\eta}}^T L^T \tilde{\bs{\lambda}}
-2\tilde{\bs{\chi}}^T \check{L}^T \tilde{\bs{\lambda}} \\
&\hspace{3.7cm}
+\bar{\bs{c}}^T\Delta_V\bs{c}
-\bs{B}^T\cdot\left(\bs{B}-2L^T\tilde{\bs{A}}\right)
\Big\}.
\end{split}
\ee
After eliminating the NL field $\bs{B}$, we get
\be
S' 
= \frac{1}{2g^2}\Big\{
\tilde{\bar{\bs{\phi}}}^T\Delta_V \tilde{\bs{\phi}}
+\bar{\bs{c}}^T\Delta_V\bs{c}
+ \tilde{\bs{A}}^T\Delta_E \tilde{\bs{A}}
-2\tilde{\bs{\eta}}^T L^T \tilde{\bs{\lambda}}
-2\tilde{\bs{\chi}}^T \check{L}^T \tilde{\bs{\lambda}}
\Big\},
\ee
where $\Delta_E = LL^T+\check{L}\check{L}^T$.

From this quadratic 1-loop effective action, we
can perform the path integral for the non-zero modes
explicitly and the 1-loop determinant becomes
\be
\begin{split}
(\text{1-loop det})&=
\frac{\det' \Delta_V}{\det' \Delta_V}
\frac{\left(\det' \Delta_V\det' \Delta_F\det' \Delta_E\right)^{1/4}}
{\sqrt{\det' \Delta_E}}\\
&=
\left(
\frac{\det' \Delta_V\det' \Delta_F} {\det' \Delta_E}
\right)^{1/4}=1\,,
\end{split}
\ee
up to an irrelevant sign factor due to the Pfaffian from the fermions, 
where
the {\it prime} denotes that the zero modes are omitted in the evaluation of the determinant 
and we have used the fact that the non-zero eigenvalues 
of $\Delta_E$ are identical with those of $\Delta_V$ and $\Delta_F$:
\be
\Spec' \Delta_V\oplus\Spec' \Delta_F
=\Spec' \Delta_E\,.
\label{eq:spectrum}
\ee

Since the zero modes are dropped from the 1-loop effective action
$S_B^{\text{1-loop}}$ and $S_F^{\text{1-loop}}$,
there exist residual integrals over the zero modes
after integrating out the non-zero modes,
namely, the vev reduces to the integral over the zero modes
\be
\left\langle f[X] \right\rangle = {\cal N}\int d\phi_0 d\bar{\phi}_0 dY_0
\left(\prod_{I=1}^{2h}d\lambda_0^{2h-I+1}\right) d\chi_0 d\eta_0\,
f[X_0]e^{\frac{1}{2g^2} Y_0^2} \,,
\label{eq:vev}
\ee
up to an irrelevant normalization constant ${\cal N}$
of the path integral measure.
Note that this reduction works only when $f[X]$ is $Q$-closed.
In particular, by setting $f[X]=1$, we again see that the partition function vanishes due to the Grassmann integral as pointed out in the previous subsection. 

\subsection{Compensating for the zero modes}
\label{compensator}

As we have discussed above, 
the vev of the operator 
reduces to the residual integral over the zero modes
after integrating out the non-zero modes. 
Therefore, so that the operator has non-trivial vev, 
we need to insert at least suitable fermion zero modes.
In the previous subsection, we simply added the mass term 
of the fermion zero modes \eqref{eq:mass term},
but such a {\it not} $Q$-invariant operator is not appropriate in the present situation. 
Instead, we here consider $Q$-closed operators
in order not to spoil the above localization argument.
Such $Q$-closed operators behave as mass terms for the fermions,
namely an exponential of the fermion bi-linear term,
including all the fermion zero modes. 
We call this kind of physical operator the compensator in the following
\cite{Kamata:2016xmu}.



First, we define a $Q$-exact operator which includes 
$\bs{\eta}$ and $\bs{\chi}$ as 
\be
{\cal O}_{\eta\chi}\equiv
Q\left(\chi_f \bar{\cal W}'(\bar{\phi}^f)\right)
=Y_f\bar{\cal W}'(\bar{\phi}^f)
+2\bar{\cal W}''(\bar{\phi}_f)\eta_f \chi^f,
\label{eq:Oetachi}
\ee
where $\bar{\cal W}'(\bar{\phi}^f)$
is an analytic function of $\bar{\phi}^f$ only\footnote{
Since we can regard $\bar{\cal W}$ as the superpotential,
the first and second derivatives of $\bar{\cal W}$ appear here.
}.
If we insert the exponential of this operator into the path integral, 
all the fields are effectively replaced by their zero modes as mentioned in
the previous subsection and the inserted operator reduces to
\begin{equation}
  e^{-\frac{1}{g^2}{\cal O}_{\eta\lambda}} \to 
-\frac{2}{g^2}\bar{\cal W}''(\bar{\phi}_0)\eta_0 \chi_0
e^{-\frac{1}{g^2} Y_0\bar{\cal W}'(\bar{\phi}_0)}
\label{eq:comp-1}
\end{equation}
in the integrand. 
It compensates for the zero modes of $\bs{\eta}$ and $\bs{\chi}$ as expected. 


We next define an operator which includes bi-linear terms
of $\bs{\lambda}$ to compensate for the zero-mode integral of $\lambda_0^I$.
As one of the candidates of the fermion bi-linear term,
we now consider an operator 
\be
\frac{1}{2}\bs{\lambda}^T \Omega \bs{\lambda},
\ee
where $\Omega$ is an anti-symmetric $n_E\times n_E$ matrix, namely $\Omega^T = -\Omega$.
The supercharge $Q$ is acting on this by
\be
\begin{split}
\frac{1}{2}Q\left( \bs{\lambda}^T \Omega \bs{\lambda}\right)
& = -\bs{\phi}^T L^T \Omega \bs{\lambda}.
\end{split}
\label{lambda-lambda transf}
\ee
On the other hand, we consider the operator piece
\be
\check{\bs{\phi}}^T \cdot \check{L}^T\bs{A},
\label{operator piece}
\ee
where $\check{\bs{\phi}}$ is a dual scalar field on the face $F$,
which is linearly constructed from the original scalar field $\bs{\phi}$
on $V$ via
\be
\check{\bs{\phi}} = M \bs{\phi},
\label{dual scalar}
\ee
where $M$ is an $n_F \times n_V$ matrix.
Note here that there are multiple candidates for $M$;
typically $\check{\phi}^f$ is determined by
$\phi^v$ on the representative point at the boundary of the face. 

Substituting (\ref{dual scalar}) into (\ref{operator piece}) and applying the $Q$-transformation, we find
\be
Q \left(\bs{\phi}^T M^T \check{L}^T \bs{A}\right) = \bs{\phi}^T M^T \check{L}^T \bs{\lambda}.
\label{BF transf}
\ee
Then, using (\ref{lambda-lambda transf}) and (\ref{BF transf}),
we see that the combined operator
\be
{\cal O}_{\lambda\lambda} = \check{\bs{\phi}}^T \check{L}^T\bs{A} 
+ \frac{1}{2}\bs{\lambda}^T \Omega \bs{\lambda},
\label{eq:Oll}
\ee
is $Q$-closed if there is a relation between the matrices
\be
\Omega L + \check{L} M=0.
\ee
So the operator ${\cal O}_{\lambda\lambda}$ could become a non-trivial observable.
We here note that the operator \eqref{eq:Oll} is $Q$-closed but is {\em not} $Q$-exact. 

As well as ${\cal O}_{\eta\chi}$, if we insert $e^{-i{\cal O}_{\lambda\lambda}}$
in the path integral, 
it reduces to an integrand of the zero modes 
as a summation over the saddle points
\begin{equation}
  e^{-i{\cal O}_{\lambda\lambda}} \to 
  \sum_{k \in {\mathbb Z}}e^{-i \left(2\pi k \phi_0 
  - \frac{1}{2}\lambda_0^I (\bs{e}_0^I)^T \Omega \bs{e}_0^J \lambda_0^J \right)}
  ={\cal C}\left(\prod_{I=1}^{2h}\lambda_0^I\right)
  \sum_{n\in {\mathbb Z}} \delta(\phi_0-n),
  \label{eq:comp-2}
\end{equation}
where ${\cal C}$ is an irrelevant constant
and we have used that $\bs{A}$ satisfies
\be
{{\check{L}^T{}}^f}_e \hat{A}^e = 2\pi k^f,
\ee
at the saddle point with the total magnetic flux $k=\sum_{f\in F}k^f$,
and the Poisson summation formula $\sum_{k\in{\mathbb Z}} e^{2\pi i kx} = \sum_{n\in{\mathbb Z}} \delta(x-n)$.
It compensates for the zero modes of $\bs{\lambda}$ as expected.

Then, if we insert the compensator 
$e^{\frac{1}{g^2}{\cal O}_{\eta\chi}}e^{-i{\cal O}_{\lambda\lambda}}$ in the path integral, 
we can compensate for all the fermion zero modes and make the integration well-defined. 
However, this is not the only effect of the compensator: 
it can handle not only fermion zero modes but also the boson zero modes at the same time.

To see this, let us consider the vev
\be
\left\langle
f[\bs{\phi}]\,
e^{-\frac{1}{2g^2}{\cal O}_{\eta\chi}}\,
e^{-i {\cal O}_{\lambda\lambda}}
\right\rangle
\ee
using the localization method,
where $f[\bs{\phi}]$ is an analytic function constructed by $\bs{\phi}$ only
and trivially $Q$-closed. 
Using \eqref{eq:vev} with \eqref{eq:comp-1} and \eqref{eq:comp-2}, 
we obtain 
\begin{align}
\left\langle f[\bs{\phi}]\,
e^{-\frac{1}{2g^2}{\cal O}_{\eta\chi}}\,
e^{-i {\cal O}_{\lambda\lambda}}
\right\rangle
&= -\frac{2}{g^2}{\cal N}{\cal C}
\int\left(\prod_{I=1}^{2h}d\lambda_0^{2h-I+1}\right)d\chi_0 d\eta_0\,
\left( \eta_0\chi_0\prod_{I=1}^{2h}\lambda_0^I \right) \nn \\
&\qquad \times \int d\phi_0 d\bar{\phi}_0dY_0 \,
\bar{\cal W}''(\bar{\phi}_0) f(\phi_0)
e^{\frac{1}{2g^2} (Y_0^2 - 2\bar{\cal W}'(\bar\phi_0)Y_0)}
\sum_{n\in{\mathbb Z}} \delta(\phi_0 - n)\nn \\
&= 
-\frac{\sqrt{8\pi}}{g} {\cal N}{\cal C}
\sum_{n\in {\mathbb Z}} f(n)\,,
\end{align}
where, after eliminating $Y_0$, 
the integration over $\bar\phi_0$ becomes just a Gaussian integral 
by changing the variable from $\bar\phi_0$ to $\bar{\cal W}'(\bar\phi_0)$. 
This result of course holds for $f[X]=1$, 
and thus the partition function is also regularized by inserting 
the compensators%
\footnote{
We also have to regularize $\sum_{n\in {\mathbb Z}}1$ in some way though. 
}. 
In this sense, the compensators give a regularization of
the zero modes in the Abelian theory.

\section{Non-Abelian Gauge Theory}
\label{Non-Abelian Theory}

In this section, we generalize our discussion to the non-Abelian gauge group $G=U(N)$.
The bosonic variables (fields) on $V$, $E$, and $F$ are denoted by $\Phi^v$, $U^e$, and $Y^f$, respectively.
Also, the fermionic fields $\eta^v$, $\lambda^e$, and $\chi^f$ exist on $V$, $E$, and $F$, respectively.
For the non-Abelian gauge group, all the fields except for $U^e$ 
are extended to the adjoint representation of $U(N)$, namely $N\times N$ matrices,
while the edge variables $U^e$ are $N\times N$ unitary matrices. 

The supersymmetry transformation is given by 
\be
\begin{array}{lcl}
Q \Phi^v = 0, && \\
Q\bar{\Phi}^v = 2\eta^v, && Q \eta^v =\frac{i}{2}[\Phi^v,\bar{\Phi}^v],\\
QU^e=i\lambda^e U^e , && Q \lambda^e = -{{L_U}^e}_v \Phi^v+i\lambda^e\lambda^e,\\
Q Y^f = i[\check{\Phi}^f, \chi^f], && Q \chi^f = Y^f,
\end{array}
\ee
where $L_U$ is defined as a gauge covariant incidence matrix;
\be
{{L_U}^e}_v \Phi^v \equiv U^e \Phi^{t(e)}{U^e}^{\dag}- \Phi^{s(e)},
\ee
and $f$ of $\check{\Phi}^f$ denotes the representative vertex of the face $f$.

It is sometimes useful to define 
\be 
\Lambda^e \equiv \lambda^e U^e\,.
\ee
The supersymmetry transformations for $U^e$ and $\Lambda^e$ become
\be
\begin{split}
QU^e = i\Lambda^e, \quad 
Q\Lambda^e 
= - (U^e \Phi^{t(e)}-\Phi^{s(e)}U^e).
\end{split}
\ee

We can construct the supersymmetric action in 
the $Q$-exact form;
\be
S = -\frac{1}{2g^2}  Q\Tr 
 \left\{
 \frac{i}{2}\eta_v [\Phi^v,\bar{\Phi}^v]
 +(\bar{\Phi}_{t(e)}U_e^\dag - U_e^\dag\bar{\Phi}_{s(e)}) \Lambda^e 
 + \chi_f\left(Y^f-2\mu(P^f)\right)
\right\}.
\label{non-Abelian action}
\ee
The moment map $\mu(P^f)$ in the action is a function of the plaquette variable on each face labeled by $f$,
which is defined by an ordered product around a face
\be
P^f \equiv \left(U^{e_1}\right)^{{{\check{L}^T{}}^f}_{e_1}} \left(U^{e_2}\right)^{{{\check{L}^T{}}^f}_{e_2}} \cdots \left(U^{e_n}\right)^{{{\check{L}^T{}}^f}_{e_n}},
\ee
where $\{e_1,e_2, \cdots, e_n\}$ are edges that surround the face $f$ in this order.
We choose the function $\mu(P^f)$ 
so that it has a unique vacuum at $P^f=1$ and asymptotically behaves 
as the field strength of the gauge field around the vacuum in order
to induce the gauge kinetic term 
after eliminating the auxiliary field $Y^f$ \cite{Sugino:2004qd,Matsuura:2014pua}.

Using the supersymmetry transformations, the bosonic part becomes
\be
S_B = \frac{1}{2g^2} \Tr 
 \left\{
 \frac{1}{4}[\Phi^v,\bar{\Phi}^v]^2
 + |U^e \Phi^{t(e)}-\Phi^{s(e)}U^e|^2
 - Y_f\left(Y^f-2\mu(P^f)\right)
\right\},
\ee
and the fermionic part becomes
\begin{multline}
S_F = -\frac{1}{2g^2} \Tr 
 \Big\{
- i\eta_v[\Phi^v,\eta^v]
 + 2(\eta_{t(e)}U_e^\dag-U_e^\dag\eta_{s(e)})\Lambda^e\\
  -i(\bar{\Phi}_{t(e)}U_e^\dag\Lambda_eU_e^\dag - U_e^\dag\Lambda_e U_e^\dag\bar{\Phi}_{s(e)}) \Lambda^e\\
-i\chi_f[\check{\Phi}^f,\chi^f]
+2\chi_f Q\mu(P^f)
\Big\}.
\end{multline}

Using the $Q$-exact action, the partition function of the non-Abelian gauge
theory is given by
\be
Z = \int \prod_{v\in V} \D \Phi^v \D \bar{\Phi}^v \D \eta^v 
\prod_{e\in E} \D U^e \D \lambda^e
\prod_{f\in F} \D Y^f \D \chi^f \,
e^{-S}.
\ee
Under the $U(1)_A$ rotation, each field transforms as
\be
\begin{split}
&\Phi^v\to e^{2i\theta_A}\Phi^v,\quad
\bar{\Phi}^v\to e^{-2i\theta_A}\bar{\Phi}^v,\quad
U^e \to U^e,\quad
Y^f \to Y^f,\\
&\eta^v\to e^{-i\theta_A}\eta^v,\quad
\lambda^e\to e^{i\theta_A}\lambda^e,\quad
\chi^f\to e^{-i\theta_A}\chi^f.
\end{split}
\ee
Then, the path integral measure of the fermions has a $U(1)_A$ anomaly
\begin{multline}
\prod_{v\in V} \D \eta^v 
\prod_{e\in E} \D \lambda^e
\prod_{f\in F} \D \chi^f\\
\to
e^{i\dim U(N) \times(n_V-n_E+n_F) \theta_A}
\prod_{v\in V} \D \eta^v 
\prod_{e\in E} \D \lambda^e
\prod_{f\in F} \D \chi^f\\
=
e^{iN^2 \chi_h \theta_A}
\prod_{v\in V} \D \eta^v 
\prod_{e\in E} \D \lambda^e
\prod_{f\in F} \D \chi^f.
\end{multline}
We will later see that this $U(1)_A$ anomaly essentially comes from
the fermion zero modes.

As usual localization arguments, we can show that the $Q$-exact action (\ref{non-Abelian action})
is independent of an overall coupling constant $t$ of the rescaled action $S\to tS$.
So the saddle point approximation in the limit of
$t\to \infty$ becomes exact
and the path integral is localized at the saddle (fixed) points.
Form the bosonic part of the action $S_B$, we find that the saddle points (localization fixed points) are given by the
equations
\bea
[\Phi^v,\bar{\Phi}^v] &=& 0,\label{diagonal condition}\\
{{L_U}^e}_v\Phi^v&=&0\label{perm condition},\\
\mu(P^f) &=& 0\label{moment map condition}.
\eea

The first equation (\ref{diagonal condition}) shows that $\Phi^v$ are diagonal.
We denote this diagonal solution by
\be
\hat{\Phi}^v=\diag(\hat{\phi}^{v,1},\hat{\phi}^{v,2},\ldots,\hat{\phi}^{v,N}).
\ee
Using this diagonal expression, we can solve the second equation (\ref{perm condition}) by
\bea
&& \hat{\phi}^{t(e),\pi_e(i)} = \hat{\phi}^{s(e),i},\label{permutation}\\
&&U^e = \hat{U}^e \Pi^e,
\eea
where $\hat{U}^e$ is a diagonal matrix,
\begin{equation}
\hat{U}^e=\diag(\hat{U}^{e,1},\hat{U}^{e,2},\cdots,\hat{U}^{e,N})\,,
\quad \left(|\hat{U}^{e,i}|=1\right)
\end{equation}
and $\Pi^e$ is a permutation matrix which represents the order $N$ permutation $\pi_e \in \mathfrak{S}_N$ on the edge $e$.

Since the permutation belongs to the Weyl group of $U(N)$, 
we can choose $\Pi^e=1$ without loss of generality
by using a gauge transformation.
So the diagonal element of $\Phi^v$ on each vertex is written by a common diagonal element
independent of $v$, that is,
\be
\hat{\phi}^{v,i}= \phi_0^i,
\ee
which stands for a ``constant'' zero mode.

Finally, Eq.~(\ref{moment map condition}) implies a constraint on $U^{e,i}$; 
\be
\prod_{e\in f}\left(\hat{U}^{e,i}\right)^{{{\check{L}^T}{}^f}_{e}}=1,
\label{eq:Uf=1}
\ee
for each $f$ and $i$.

Now let us consider an effective action near the saddle point.
We expand $\Phi^v$ and $U^e$
around the solution to the saddle point equation as
\be
\begin{split}
\Phi^v &= \hat{\Phi}+ \frac{1}{\sqrt{t}}\tilde{\Phi}^v,\\
U^e &=  e^{\frac{i}{\sqrt{t}}\tilde{A}^e}\hat{U}^e \simeq \left(1+\frac{i}{\sqrt{t}} \tilde{A}^e\right)\hat{U}^e,
\end{split}
\ee
where $\hat{\Phi}=\diag(\phi_0^1,\phi_0^2,\ldots,\phi_0^N)$.
All other fields including the fermions are treated as the fluctuations and rescaled by $1/\sqrt{t}$
and
we omit {\it tilde} for these fluctuations.
Using the Cartan-Weyl basis (see Appendix \ref{Cartan-Weyl}), the Cartan parts are written as
\be
\hat{\Phi} = \phi_0^i\mathtt{H}_i,\quad
\hat{\bar{\Phi}} = \bar{\phi}_0^i\mathtt{H}_i,\quad
\hat{U}^e = \hat{U}^{e,i}\mathtt{H}_i,
\ee
and the fluctuations and fermions are expanded as follows;
\be
\begin{split}
&\tilde{\Phi}^v = \tilde{\phi}^{v,i}\mathtt{H}_i+ \tilde{\phi}^{v,\alpha}\mathtt{E}_\alpha,\quad
\tilde{\bar{\Phi}}^v = \tilde{\bar{\phi}}^{v,i}\mathtt{H}_i+ \tilde{\bar{\phi}}^{v,\alpha}\mathtt{E}_\alpha,\quad
\tilde{A}^e = \tilde{A}^{e,i}\mathtt{H}_i+ \tilde{A}^{e,\alpha}\mathtt{E}_\alpha,\\
&\eta^v = \eta^{v,i}\mathtt{H}_i+ \eta^{v,\alpha}\mathtt{E}_\alpha,\quad
\chi^f = \chi^{f,i}\mathtt{H}_i+ \chi^{f,\alpha}\mathtt{E}_\alpha,\quad
\lambda^e = \lambda^{e,i}\mathtt{H}_i+ \lambda^{e,\alpha}\mathtt{E}_\alpha,
\end{split}
\ee
where the upper and lower indices of $i$ and $\alpha$ are contracted.

Using these expansions, we find
\be
\begin{split}
[\Phi^v,\bar{\Phi}^v] &= \frac{1}{\sqrt{t}}\left(
[\hat{\Phi},\tilde{\bar{\Phi}}^v]+[\tilde{\Phi}^v,\hat{\bar{\Phi}}]
\right) + {\cal O}(1/t),\\
&=\frac{1}{\sqrt{t}}\left(
\alpha(\phi_0)\tilde{\bar{\phi}}^{v,\alpha}
-\alpha(\bar{\phi}_0)\tilde{\phi}^{v,\alpha}
\right)\mathtt{E}_\alpha + {\cal O}(1/t),\\
{{L_U}^e}_v \Phi^v
 &=\frac{1}{\sqrt{t}}\left(
 {L^e}_v\tilde{\phi}^{v,i}\mathtt{H}_i
 +
{{L_{\hat{U}}}^e}_v\tilde{\phi}^{v,\alpha}\mathtt{E}_\alpha
-i\alpha(\phi_0)\tilde{A}^{e,\alpha}\mathtt{E}_\alpha
\right)+{\cal O}(1/t),\\
\mu(P^f)
&=\frac{1}{\sqrt{t}}\left(
{{\check{L}^T{}}^f}_e \tilde{A}^{e,i}\mathtt{H}_i
+{{\check{L}^\dag_{\hat{U}}{}}^f}_e\tilde{A}^{e,\alpha}\mathtt{E}_\alpha
 \right)+{\cal O}(1/t),
\end{split}
\ee
up to the leading order, where
\be
\alpha(\phi_0) \equiv \alpha_i \phi_0^i,\quad
\alpha(\bar{\phi}_0) \equiv \alpha_i \bar{\phi}_0^i.
\ee
When we expand the moment map $\mu(P^f)$, we require the covariant dual incidence matrix
${{\check{L}^\dag_{\hat{U}}{}}^f}_e$,
which is defined by
\be
\left.\delta \mu(P^f)\right|_{U^e=\hat{U}^e} = 
i\sum_{e \in f}{{\check{L}^\dag_{\hat{U}}{}}^f}_e \delta A^e
\equiv i\sum_{e \in f}{{\check{L}^\dag{}}^f}_e {\hat{X}^f}_e \delta A^e {\hat{X}^{\dag f}}_e,
\ee
with
\be
{\hat{X}^f}_{e_i}=
\begin{cases}
\hat{U}^{\check{L}^f_{e_1}}_{e_1}\hat{U}^{\check{L}^f_{e_2}}_{e_2}\cdots\hat{U}^{\check{L}^f_{e_{i-1}}}_{e_{i-1}}
& \text{if }{\check{L}^f{}}_{e_i}=+1\\
\hat{U}^{\check{L}^f_{e_1}}_{e_1}\hat{U}^{\check{L}^f_{e_2}}_{e_2}\cdots\hat{U}^{\check{L}^f_{e_i}}_{e_i}
& \text{if }{\check{L}^f{}}_{e_i}=-1
\end{cases}.
\ee

Using the above expansions up to quadratic order of fluctuations, we obtain the rescaled 1-loop
effective bosonic action;
\begin{align}
S_B^{\text{1-loop}}\equiv &\lim_{t\to \infty}t S_B \nn \\
= &\frac{1}{2g^2}\Bigg[\sum_{i=1}^N\Big\{
|{L^e{}}_v\tilde{\phi}^{v,i}|^2
-Y^i_f(Y^{f,i}-2{{\check{L}^T{}}^f}_e\tilde{A}^{e,i})
\Big\} \nn \\
&\hspace{5mm} +\sum_{\alpha}\Big\{
\frac{1}{4}\left|\alpha(\phi_0)\tilde{\bar{\phi}}^{v,\alpha}
-\alpha(\bar{\phi}_0)\tilde{\phi}^{v,\alpha}\right|^2 
+|{{L_{\hat{U}}}^e{}}_v\tilde{\phi}^{v,\alpha}|^2
+|\alpha(\phi_0)\tilde{A}^{e,\alpha}|^2 \nn \\
&\hspace{2cm}  -Y^{-\alpha}_f(Y^{f,\alpha}-2{{\check{L}^\dag_{\hat{U}}{}}^f}_e\tilde{A}^{e,\alpha})
\Big\}
\Bigg].
\label{Sb-1loop}
\end{align}
Similarly, the fermionic part of the 1-loop effective action becomes 
\begin{align}
S_F^{\text{1-loop}}\equiv &\lim_{t\to \infty}t S_F\ \nn \\
= &-\frac{1}{2g^2}\Bigg[\sum_{i=1}^N\Big\{
2\eta^i_v{{L^T{}}^v}_e\lambda^{e,i}
+2\chi^{i}_f{{\check{L}^T{}}^f}_e\lambda^{e,i}
\Big\} \nn \\
&\hspace{8mm} +\sum_{\alpha}\Big\{
2\eta^{-\alpha}_v{{L_{\hat{U}}^\dag{}}^v}_e\lambda^{e,\alpha}
+2\chi^{-\alpha}_f{{\check{L}_{\hat{U}}^\dag{}}^f}_e\lambda^{e,\alpha} \nn \\
&\hspace{2cm} -i\alpha(\phi_0)\eta^{-\alpha}_v \eta^{v,\alpha}
-i\alpha(\phi_0)\chi^{-\alpha}_f \chi^{f,\alpha}
+i\alpha(\bar{\phi}_0)\lambda^{-\alpha}_e\lambda^{e,\alpha}
\Big\}
\Bigg].
\label{Sf-1loop}
\end{align}
This effective action gives the same path integral as the original action owing to the $Q$-exactness of the action.

We first integrate over the components of the root vectors.
To this end, 
we fix the gauge symmetry $U(1)^N$ in the 1-loop actions \eqref{Sb-1loop} and \eqref{Sf-1loop} 
by introducing the FP ghost $(c^{v,\alpha},\bar{c}^{v,\alpha})$ and 
NL field $B^{v,\alpha}$. 
The corresponding BRST transformations are given by
\be
\begin{split}
&\delta_B c^{v,\alpha} = 0,\quad
\delta_B \bar{c}^{v,\alpha} = 2 B^{v,\alpha},\quad
\delta_B B^{v,\alpha} = 0,\\
&\delta_B \tilde{\phi}^{v,\alpha} = -i\alpha(\phi_0)c^{v,\alpha},\quad
\delta_B \tilde{\bar{\phi}}^{v,\alpha} = -i\alpha(\bar{\phi}_0)c^{v,\alpha}, \\
&\delta_B \tilde{A}^{e,\alpha} = - {{L_{\hat{U}}}^e}_v c^{v,\alpha},
\end{split}
\ee
where we assume that the FP ghost and NL field are the same order in $t$
as the fluctuations.
We define the gauge fixing function for the root vectors by
\be
f^{v,\alpha} \equiv B^{v,\alpha} -2{{L_{\hat{U}}^\dag{}}^v}_e\tilde{A}^{e,\alpha}
+i \alpha(\phi_0)\tilde{\bar{\phi}}^{v,\alpha}
+i \alpha(\bar{\phi}_0)\tilde{\phi}^{v,\alpha}.
\ee
Then the gauge fixing and FP ghost term is given in the BRST exact form by
\be
\begin{split}
S^{\text{root}}_{\text{GF+FP}} &= -\frac{1}{4g^2} \delta_B \sum_{\alpha}\bar{c}_v^{-\alpha}f^{v,\alpha}\\
&=\frac{1}{2g^2}\sum_{\alpha}\Bigg[
-B_v^{-\alpha}f^{v,\alpha}
+\bar{c}_v^{-\alpha}\left(
{{\Delta^V_{\hat{U}}{}}^v}_{v'}
+ |\alpha(\phi_0)|^2{\delta^v}_{v'}
\right)c^{v',\alpha}\Bigg],
\end{split}
\label{eq:ghost action}
\ee
where
\be
{{\Delta^V_{\hat{U}}{}}^v}_{v'}\equiv
{{L_{\hat{U}}^\dag{}}^v}_e{{L_{\hat{U}}{}}^e}_{v'}.
\ee
After eliminating the NL filed $B^{v,\alpha}$, we get the action for
$(\tilde{\phi}^{v,\alpha},\tilde{\bar{\phi}}^{v,\alpha})$
\be
S^{\text{root}}_{(\tilde{\phi},\tilde{\bar{\phi}})}
=\frac{1}{2g^2}
\sum_\alpha
\tilde{\bar{\phi}}_v^{-\alpha}\left(
{{\Delta^V_{\hat{U}}{}}^v}_{v'}
+ |\alpha(\phi_0)|^2{\delta^v}_{v'}
\right)\tilde{\phi}^{v',\alpha},
\ee
whose 1-loop determinant is completely canceled with
the contribution from the ghost part in \eqref{eq:ghost action}. 

In addition, integrating out the auxiliary field $Y^{f,\alpha}$,
the action for the gauge boson reduces to
\be
S^{\text{root}}_{\tilde{A}}
=\frac{1}{2g^2}
\sum_\alpha
\tilde{A}_e^{-\alpha}
\left(
{{\Delta^E_{\hat{U}}{}}^e}_{e'}
+|\alpha(\phi_0)|^2{\delta^e}_{e'}
\right)
\tilde{A}^{e',\alpha},
\ee
where
\be
{{\Delta^E_{\hat{U}}{}}^e}_{e'}
\equiv
{{L_{\hat{U}}{}}^e}_{v}{{L_{\hat{U}}^\dag{}}^v}_{e'}
+{{\check{L}_{\hat{U}}{}}^e}_{v}{{\check{L}_{\hat{U}}^\dag{}}^v}_{e'}.
\ee
Then we obtain the 1-loop determinant
for the gauge boson $\tilde{A}^{e,\alpha}$
\be
\prod_{\alpha>0}
\frac{1}{|\alpha(\phi_0)|^{2n_E^0}\det'\left(
\Delta^E_{\hat{U}}+|\alpha(\phi_0)|^2
\right)},
\label{boson 1-loop}
\ee
where $n_E^0$ is the number of the zero eigenstates
for the edge Laplacian
and
the {\it prime} on the determinant stands for omitting the zero eigenvalues.

Next, let us consider the integral of the fermions.
We need to care about the Laplacian zero modes for the fermions,
but we get the 1-loop determinant
\begin{multline}
\prod_{\alpha>0}
\alpha(\phi_0)^{n^0_V+n^0_F}
\alpha(\bar{\phi}_0)^{n_E^0}\\
\times
\sqrt{
{\det}'\left(
\Delta^V_{\hat{U}}+|\alpha(\phi_0)|^2
\right)
{\det}'\left(
\Delta^F_{\hat{U}}+|\alpha(\phi_0)|^2
\right)
{\det}'\left(
\Delta^E_{\hat{U}}+|\alpha(\phi_0)|^2
\right)
}\\
=\prod_{\alpha>0}
\alpha(\phi_0)^{n^0_V+n^0_F}
\alpha(\bar{\phi}_0)^{n_E^0}
{\det}'\left(
\Delta^E_{\hat{U}}+|\alpha(\phi_0)|^2
\right),
\label{fermion 1-loop}
\end{multline}
where we have used that
the non-zero eigenvalues of $\Delta^E_{\hat{U}}$
is a combination of the non-zero eigenvalues of $\Delta^V_{\hat{U}}$ and
$\Delta^F_{\hat{U}}$, namely\footnote{
We can see that the condition \eqref{eq:Uf=1} guarantees the orthogonality 
like (\ref{orthogonality}) for $L_{\hat{U}}$ and $\check{L}_{\hat{U}}$ 
in the concrete examples. See Appendix A.
} 
\be
\Spec'\Delta^V_{\hat{U}} \oplus \Spec'\Delta^F_{\hat{U}}
=\Spec'\Delta^E_{\hat{U}}.
\ee

Combining the 1-loop determinant of the bosons (\ref{boson 1-loop})
and fermions (\ref{fermion 1-loop}),
we finally obtain the total 1-loop determinant for the root vector
components
\be
\prod_{\alpha>0}
\alpha(\phi_0)^{n^0_V+n^0_F-n^0_E}
=\prod_{\alpha>0}\alpha(\phi_0)^{\chi_h},
\ee
where we have used the index theorem on the graph Laplacians
as well as the Abelian theory.
Note here that the above 1-loop determinant has an anomaly phase under
$U(1)_A$ symmetry as
\be
\prod_{\alpha>0}\alpha(\phi_0)^{\chi_h}
\to
e^{iN(N-1)\chi_h \theta_A}\prod_{\alpha>0}\alpha(\phi_0)^{\chi_h}.
\ee

Let us next think about the Cartan part of the fluctuations and fermions.
The Cartan part is nothing but $N$ copies of the Abelian theory
discussed in Sec.~\ref{Localization in the Abelian theory}.
Introducing the FP ghost $(c^{v,i},\bar{c}^{v,i})$,
NL field $B^{v,i}$ and gauge fixing function for the Cartan modes, 
\be
f^{v,i} = B^{v,i}-2{{L^T}^v}_e\tilde{A}^{e,i},
\ee
the gauge fixing term for the Cartan modes is given by
\be
\begin{split}
S^{\text{Cartan}}_{\text{GF+FP}} &= -\frac{1}{4g^2} \delta_B \sum_{i=1}^N\bar{c}_v^i
f^{v,i}\\
&=\frac{1}{2g^2}\sum_{i=1}^N\Bigg[
-B_v^{i}f^{v,i}
+\bar{c}_v^i
{{\Delta_V{}}^v}_{v'}
c^{v',i}\Bigg].
\end{split}
\ee
We find that the 1-loop determinant of the bosons $(\tilde{\phi}^{v,i},\tilde{\bar{\phi}}^{v,i})$
and $(c^{v,i},\bar{c}^{v,i})$ are canceled with each other.
After eliminating the auxiliary field $Y^{f,i}$ and NL field $B^{v,i}$,
the integral of the gauge boson $\tilde{A}^{e,i}$ gives the 1-loop determinant
\be
\frac{1}{\left({\det}' \Delta_E\right)^N}.
\ee
This 1-loop determinant for the bosons is canceled with
a 1-loop determinant for the fermions
\be
\left({\det}' \Delta_V {\det}' \Delta_F {\det}' \Delta_E\right)^{N/2},
\ee
by using the same fact as Eq.~\eqref{eq:spectrum} for the Cartan part. 

Thus we finally obtain the path integral measure over the zero modes
in the Cartan subalgebra
\be
Z = {\cal N}\int \prod_{i=1}^N
d\phi_0^i d\bar{\phi}^i_0 dY_0^i
\left(\prod_{I=1}^{2h} d\lambda_0^{i,2h-I+1}\right)
d\chi_0^i 
d\eta_0^i\, 
e^{\frac{1}{2g^2} (Y_0^{i})^2 }
\prod_{\alpha>0}\alpha(\phi_0)^{\chi_h},
\label{eq:non-abelian partition function}
\ee
up to a normalization constant ${\cal N}$.
The zero-mode integral is a multiple of the Abelian gauge theory except for
the Vandermonde type determinant $\prod_{\alpha>0}\alpha(\phi_0)^{\chi_h}$.
This phenomenon is the same as what occurs in the  
continuum field theory localization
and is called ``diagonalization'' or ``Abelianization''.
(See for review \cite{Blau:1995rs}.)
This integral measure has the $U(1)_A$ anomaly
\begin{multline}
\prod_{i=1}^N
d\phi_0^i d\bar{\phi}^i_0  \left(\prod_{I=1}^{2h} d\lambda_0^{i,2h-I+1}\right)d\chi_0^id\eta_0^i \,
\prod_{\alpha>0}\alpha(\phi_0)^{\chi_h}\\
\to
e^{i N^2 \chi_h \theta_A}
\prod_{i=1}^N
d\phi_0^i d\bar{\phi}^i_0
\left(\prod_{I=1}^{2h} d\lambda_0^{i,2h-I+1}\right)
d\chi_0^i d\eta_0^i  \,
\prod_{\alpha>0}\alpha(\phi_0)^{\chi_h},
\end{multline}
as expected.

Due to the existence of the fermion zero modes,
the partition function itself is ill-defined.
So we need to insert an operator 
which compensates for the fermionic
zero modes.
As mentioned above, the path integral reduces to the multiple integrals
of the Abelian ones. So we can compensate the fermionic zero modes
by inserting the compensator discussed in Sec.~\ref{compensator}
for each Cartan part.
It however seems to be difficult to construct a compensator,
which is invariant under the non-Abelian gauge group and supersymmetric ($Q$-closed),
prior to the Abelianization.
We leave the construction of the complete compensator in the non-Abelian gauge theory
as a future problem.

\section{Conclusion and Discussion}
\label{Conclusion and Discussion}

In this paper, the properties of the discretized two-dimensional supersymmetric gauge theory (the generalized Sugino model) given in \cite{Matsuura:2014kha} were studied analytically by using the techniques of graph theory.

From a graph theory point of view, 
the model is defined on a two-dimensional graph and its dual graph, 
and the action can be efficiently described using the so-called incidence matrix $L$ and the dual incidence matrix $\check{L}$.

The incidence and the dual incidence matrix map 
from a vector on the vertices to a vector on the edges 
and form a vector on the faces to a vector on the edges, respectively, 
and obey the property such that $L\check{L}^T=0$. 
Therefore, if we consider the vectors on the vertices, edges, and faces 
as analogs of 0-form, 1-form, and 2-form, respectively, 
we can regard $L$ and ${\check{L}}^T$ as the exterior
derivatives, and $L^T$ and ${\check{L}}$ as its dual. 
The cohomology can be defined using $L$ and $\check{L}$, 
and a parallel argument of Hodge's theorem on the Riemann surfaces can be developed on the graphs. 
In particular, we found that the structure of 
the kernel of $L$ and $\check{L}$ is completely determined by the topology of the graph.

We used the properties of these matrices to examine 
the generalized Sugino model with gauge group $U(1)$ 
and found that the number of fermion zero modes depends 
on the topology of the graph. 
Since these zero modes make the partition function ill-defined, 
it is necessary to insert an appropriate operator
including zero modes in the background. 
We proposed a mass term for the fermions 
so that this operation is done automatically. 
We confirmed that the introduction of this mass term regularizes the Dirac matrix 
and makes the theory itself well-defined. 
We also derived anomalous WT identities on the graph
corresponding to the classical global $U(1)$ symmetry, 
which is broken by quantum mechanically
unless the topology of the graph is the torus. 
In the continuous theory, this anomaly appears as the scalar curvature in the WT identity. 
On the other hand, in the theory on a graph, 
a quantity related to the degrees of the vertex and face arises instead of the scalar curvature.  
This corresponds to the fact that the scalar curvature 
of a two-dimensional surface is given by the deficit angle.

We examined the generalized Sugino model from the viewpoint of topological field theory by restricting the physical quantity to $Q$-cohomology. 
We used the so-called localization technique 
to compute the expected value of a general $Q$-closed operator.
As a result of localization, the vev can be expressed in terms of the usual integration by the zero modes. 
In that case, unless the operator contains all fermion zero modes, 
the vev trivially vanishes. 
We constructed $Q$-invariant operators (compensators) 
that cancel out the fermion zero modes, 
and gave a prescription for computing the vev for nontrivial values.
The compensators introduced here regularize the theory properly. 


We also extended the graph-theoretic description to the non-Abelian theory.
Reflecting the non-commutativity of gauge groups, 
the incidence and dual incidence matrices are transformed 
to be covariant differences instead of ordinary differences. 
This transformation eliminates the orthogonality of the incidence matrix and the dual incidence matrix unless all the plaquette variables are unity ($P^f=1$) 
and the fermion zero modes that appeared in the Abelian theory are lifted
in most configurations. 
Therefore, in most configurations, the Dirac matrix is regular and the inverse exists.
However, the situation is different around 
the saddle points of the $Q$-transformation.
Using the localization technique with an appropriate gauge fixing, 
the non-Abelian generalized Sugino model can be effectively reduced to an Abelian theory. 
As a result, the evaluation of the partition function is completely parallel 
to the calculation of that of the Abelian theory, 
and the fermion zero modes arising at the saddle points make an important contribution. 
In particular, it is confirmed that the partition function becomes ill-defined 
unless these fermion zero modes on the saddle points are properly treated.

The fact that the non-Abelian generalized Sugino model 
is also affected by the fermion zero modes 
is quite important when carrying out numerical simulations. 
As mentioned above, the fermion zero modes appear only 
on the saddle points of the $Q$-transformation 
and thus the Dirac matrix is regular in almost all configurations.
Therefore the numerical simulation proceeds even without any special treatment for the zero modes.
However, since the saddle points of the $Q$-transformation are a part of classical configurations, 
the zero modes would affect the computation especially in the region close to the continuum limit, 
and there is a possibility that reliable results cannot be obtained. 
This conclusion holds even in the case of a torus background where the anomaly is canceled 
because the fermion zero modes still exist at the saddle points.

In the case of the torus, 
the fermion zero modes are completely lifted up by imposing the anti-periodic boundary condition 
in the temporal direction to the fermions. 
Therefore the numerical simulations for the system with finite temperature are expected to work well. 
All the simulations imposing the anti-periodic boundary conditions for the fermions 
in the temporal direction have been successfully carried out
\cite{Kanamori:2007ye,Kanamori:2007yx,Kanamori:2008bk,Kanamori:2008yy,Kanamori:2009dk,Hanada:2009hq}.

On the other hand, 
it was reported in \cite{Kanamori:2008bk} that numerical calculations 
did not yield the expected results for two-point functions in the continuum limit%
\footnote{
See also \cite{Kadoh:2009rw} where the WT identity is analytically examined by using a "semi-perturbative" treatment. 
}, 
while it was reported in \cite{Hanada:2009hq} that the vevs of Yukawa terms 
consistently degenerate in the continuum limit. 

At first glance, it seems that the simulation will not work 
of the presence of the fermion zero modes, 
but the situation is slightly more complicated. 
The point is that the theory is expected to have (at least) two phases; 
the phase where the eigenvalues of the scalar field form a bound state 
and the phase where they run freely \cite{Hanada:2009hq}. 
In the numerical simulation, 
the flat directions of the scalar field are controlled by introducing a mass term, 
and thus the configurations are all in the phase with the bound state. 
On the other hand, the partition function \eqref{eq:non-abelian partition function} 
is obtained by integrating out all the configurations including both phases. 
Therefore, although it is one of the possibilities, 
if the fermion zero modes are effectively lifted up in the phase with the bound state, 
the simulation would work well even if one takes the periodic boundary condition for the fermions.
It will be interesting to check whether the situation changes or not 
if we deal with the fermion zero modes in an appropriate way.

It is only when the background is torus 
that the boundary condition can eliminate all the fermion zero modes.  
This is because changing the boundary condition is equivalent to transforming $D=(L,\check{L})$.
In the case of the torus,  
$D$ is a square matrix, 
so all zero modes will be eliminated 
if we transform it in such a way that the zero modes of $L$ and $\check{L}$ are eliminated
Imposing the anti-periodic boundary condition is an example of this kind of modification. 
On the other hand, in non-torus cases, 
there are always zero modes no matter how much $D$ is deformed 
since the rank of the rectangular matrix $D$ is at most $\min(n_E, n_V+n_F)$. 
Therefore it is a peculiarity of the torus 
that the zero modes can be dealt with just by considering 
the finite temperature.

In the non-toric cases, 
we have to insert some corrections to the diagonal blocks of $\slashed{D}$ to eliminate all the fermion zero modes. 
This is equivalent to introducing a mass term to the fermions 
and the \eqref{eq:mass term} is an example of this kind of modification. 
In the case of non-Abelian theories, 
it is necessary to introduce such a mass term 
that appropriately lifts the zero modes arising on the $Q$-fixed points
without breaking the gauge symmetry and with respecting the $Q$-symmetry if possible. 
For the zero modes of $\eta$ and $\chi$, 
we can simply extend the compensator \eqref{eq:Oetachi} as 
\begin{equation}
  Q{\rm tr}\left( \chi_f \bar{\cal W}'(\bar\phi^f) \right)\,.
\end{equation}
For the zero modes of $\lambda$, however, 
it is still an open problem to construct such an operator that is $Q$-closed (not $Q$-exact) 
and includes bi-linear terms of $\lambda$ like \eqref{eq:Oll}. 
Instead, $Q$-exact operators like
\begin{equation}
  Q{\rm tr}\left(\lambda_l P_f\right)
\end{equation}
may work. 
It is also interesting to analyze the property of non-Abelian compensators.

One problem that has not yet been achieved in previous studies is the introduction of matter fields. In supersymmetric gauge theories, in order to introduce the matter field as a chiral multiplet, it is essential to consider chiral fermions.

The chiral fermions have been constructed on the regular square lattice
by various methods, but how to construct chiral fermions on a discrete space arbitrarily partitioned by a graph is completely unknown.
However, in this paper, it was clarified that the incidence matrix in graph theory has a deep connection with the Dirac operator, so it seems possible to define chiral fermions using graph theory. We could use graph theory to introduce chiral fermions on discrete spaces, to construct supersymmetric gauge theories including matter fields, and to analyze and understand chiral anomalies induced by chiral fermions. These are also important future issues.

Once the introduction of the matter field is achieved,
the interaction with the gauge field also allows the construction of solitons
such as vortices on the discretized Riemann surface.
At present, the construction of solitons on the graph
is a novel problem. It is also very interesting to understand the non-perturbative effects by such solitons in supersymmetric gauge theories on the graph.

As mentioned in the introduction, 
the continuum limit of the generalized Sugino model 
is a topologically twisted ${\cal N}=(2,2)$ supersymmetric gauge theory, 
which is a theory on the Riemann surface with a $U(1)_A$ background field balanced with the spin connection of the background space-time
and the fermions behave as fields with integer spins. Interestingly, this property is similar to the K\"ahler-Dirac fermion. 
In fact, it is argued in \cite{Catterall:2018lkj,Butt:2021brl} that 
systems with K\"ahler-Dirac fermion have an anomaly proportional to the background Euler number as well. 
These approaches would be compatible with the lattice gravity,
which realizes gravity via a random triangulation.
For example, in the formulation given in \cite{Kawamoto:1990hk}, 
it is essential to place the gauge field on the edge of 
the triangulation and the spin connection on the dual edge. 
It is remarkable that, in this formulation, the action of
the gravity is written in terms of 
the deficit angle of the dual plaquette, 
whereas the anomaly in the local WT identities appears as the deficit angle as well. 
It will be interesting to consider lattice gravity from the viewpoint of graph theory. 
In particular, 
it is suggestive that the construction of the discretized theory with the supersymmetry 
is possible only when the spin connection and the background gauge field are properly balanced.
Through research in this direction, 
we expect to obtain new insights from graph theory 
for lattice gravity in higher dimensions \cite{Kawamoto:1990hk,Kawamoto:1999jv,Kawamoto:1999tf}.

\section*{Acknowledgments}
We would like to thank S.~Kamata and T.~Misumi
for useful discussions in the early stage of this work.
We would like to thank N.~Kawamoto
for useful discussions at Hokkaido University and for teaching us stimulating references.
S.M~ would like to thank M.~Hanada and D.~Kadoh for useful discussion. 
This work is supported in part 
by Grant-in-Aid for Scientific Research (KAKENHI) (C) Grant Number 
17K05422 (K.~O.) 
and  Grant-in-Aid for Scientific Research (KAKENHI) (C), Grant Number 20K03934 (S.~M.).
A part of this work is motivated by the results 
of numerical simulations carried out on 
Oakforest-PACS computational system provided by Tokyo University 
through the HPCI System Research Project (Project ID: hp200052).

\appendix 
\section{Examples of the Graph Data}

In this Appendix, we give concrete examples of the graph data and objects and check some properties.
We also give plaquette variables on each face and covariantized version of the (dual)
incidence matrix in non-Abelian gauge theory.

\subsection{Tetrahedron}

\begin{figure}
\begin{center}
\includegraphics[scale=0.6]{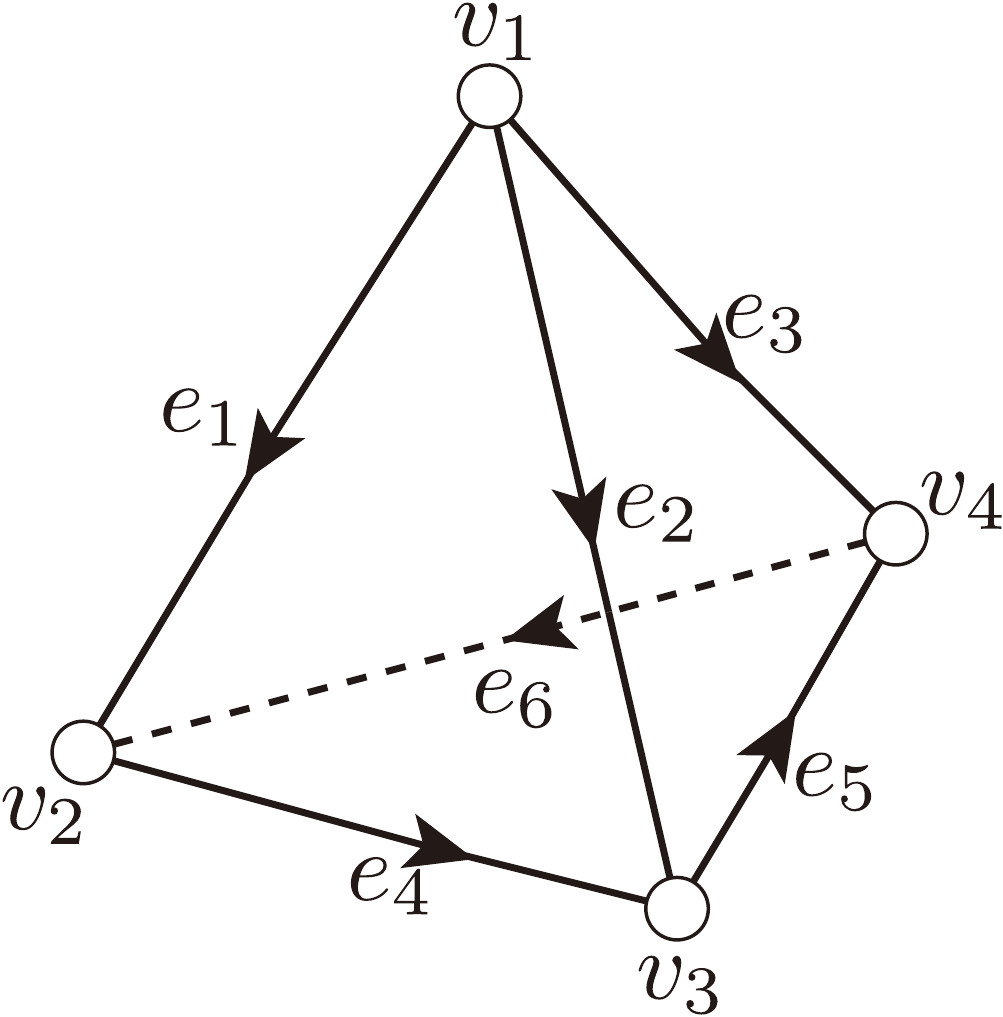}
\end{center}
\caption{A directed graph for a tetrahedron.
There are 4 vertices
and 6 directed edges.}
\label{tetrahedron graph}
\end{figure}

A directed graph associated with a tetrahedron is shown in Fig.~\ref{tetrahedron graph}.
There 4 vertices which is labeled by $V(\Gamma)=\{v_1,v_2,v_3,v_4\}$.
The directed connectivity for 6 edges is given by
$E(\Gamma)=\{e_1,e_2,e_3,e_4,e_5,e_6\}
=\{v_1\to v_2,v_1\to v_3,v_1\to v_4,v_2 \to v_3,v_3\to v_4,v_4\to v_2\}$.

For this directed graph, the incidence matrix is given by
\be
L=
\left(
\begin{array}{cccc}
 -1 & 1 & 0 & 0 \\
 -1 & 0 & 1 & 0 \\
 -1 & 0 & 0 & 1 \\
 0 & -1 & 1 & 0 \\
 0 & 0 & -1 & 1 \\
 0 & 1 & 0 & -1 \\
\end{array}
\right).
\ee
We can construct the Laplacian matrix for the vertex from the incidence matrix
\be
\Delta_V = L^T L
=\left(
\begin{array}{cccc}
 3 & -1 & -1 & -1 \\
 -1 & 3 & -1 & -1 \\
 -1 & -1 & 3 & -1 \\
 -1 & -1 & -1 & 3 \\
\end{array}
\right).
\ee
Then the adjacency matrix becomes
\be
K=\left(
\begin{array}{cccc}
 0 & 1 & 1 & 1 \\
 1 & 0 & 1 & 1 \\
 1 & 1 & 0 & 1 \\
 1 & 1 & 1 & 0 \\
\end{array}
\right).
\ee
The vertex Laplacian has the eigenvalues $\{4,4,4,0\}$,
which contain one zero.

Four faces are defined by
$f_1=\{e_1,e_4,\bar{e}_2\}$,
$f_2=\{e_2,e_5,\bar{e}_3\}$,
$f_3=\{e_3,e_6,\bar{e}_1\}$
and
$f_3=\{\bar{e}_4,\bar{e}_6,\bar{e}_5\}$.
Then the dual incidence matrix is given by
\be
\check{L}
=\left(
\begin{array}{cccc}
 1 & 0 & -1 & 0 \\
 -1 & 1 & 0 & 0 \\
 0 & -1 & 1 & 0 \\
 1 & 0 & 0 & -1 \\
 0 & 1 & 0 & -1 \\
 0 & 0 & 1 & -1 \\
\end{array}
\right).
\ee
We can see that
$L^T\check{L}=\check{L}^T L=0$.

Using the dual incidence matrix, we can construct the Laplacian for the face and edges as
\be
\begin{split}
&\Delta_F=\check{L}^T\check{L}=
\left(
\begin{array}{cccc}
 3 & -1 & -1 & -1 \\
 -1 & 3 & -1 & -1 \\
 -1 & -1 & 3 & -1 \\
 -1 & -1 & -1 & 3 \\
\end{array}
\right),\\
&\Delta_E=LL^T+\check{L}\check{L}^T
=\left(
\begin{array}{cccccc}
 4 & 0 & 0 & 0 & 0 & 0 \\
 0 & 4 & 0 & 0 & 0 & 0 \\
 0 & 0 & 4 & 0 & 0 & 0 \\
 0 & 0 & 0 & 4 & 0 & 0 \\
 0 & 0 & 0 & 0 & 4 & 0 \\
 0 & 0 & 0 & 0 & 0 & 4 \\
\end{array}
\right),
\end{split}
\ee
which have the eigenvalues $\{4,4,4,0\}$ and $\{4,4,4,4,4,4\}$, respectively.

An non-Abelian generalization for the incidence matrix acting on
the adjoint representation is given by
\be
L_U=
\left(
\begin{array}{cccc}
 -1 & U^1\cdot U^{1\dag} & 0 & 0 \\
 -1 & 0 & U^2\cdot U^{2\dag} & 0 \\
 -1 & 0 & 0 & U^3\cdot U^{3\dag} \\
 0 & -1 & U^4\cdot U^{4\dag} & 0 \\
 0 & 0 & -1 & U^5\cdot U^{5\dag} \\
 0 & U^6\cdot U^{6\dag} & 0 & -1 \\
\end{array}
\right),
\ee
where ``\ $\cdot$\ '' stands for an insertion position of matrices in the adjoint representation
when this covariant incidence matrix acts from the right;
\be
(X\cdot Y) A = XAY,
\ee
for example.

A conjugate of the dual incidence matrix is derived from 4 plaquette variables;
\be
P^1 = U^1U^4U^{2\dag},\quad
P^2 = U^2U^5U^{3\dag},\quad
P^3 = U^3U^6U^{1\dag},\quad
P^4 = U^{4\dag}U^{6\dag}U^{5\dag}.
\ee
It becomes
\be
\check{L}^\dag_U
=\begin{pmatrix}
 \cdot P^1 & - P^1\cdot & 0 & U^1\cdot U^{1\dag}P^1  &0 & 0\\
 0 & \cdot P^2 & - P^2\cdot & 0  &U^2\cdot U^{2\dag}P^2 & 0\\
 - P^3\cdot & 0 &  \cdot P^3 & 0  &0 & U^3\cdot U^{3\dag}P^3\\
 0 & 0 & 0 & -U^{4\dag}\cdot U^4P^4  &-P^4\cdot & -P^4U^5\cdot U^{5\dag}
\end{pmatrix}.
\ee
Then we find
$L_U^\dag \check{L}_U=\check{L}^\dag_U L_U=0$ iff $P^f=1$.

We can also define the covariant Laplacians by
\be
\Delta^V_U = L_U^\dag L_U, \quad
\Delta^F_U = \check{L}_U^\dag \check{L}_U, \quad
\Delta^E_U =  L_U  L_U^\dag +  \check{L}_U\check{L}_U^\dag,
\ee
which have the same eigenvalues as $\Delta_V$, $\Delta_F$ and $\Delta_E$ iff $P^f=1$.

\subsection{Torus}

\begin{figure}
\begin{center}
\includegraphics[scale=0.6]{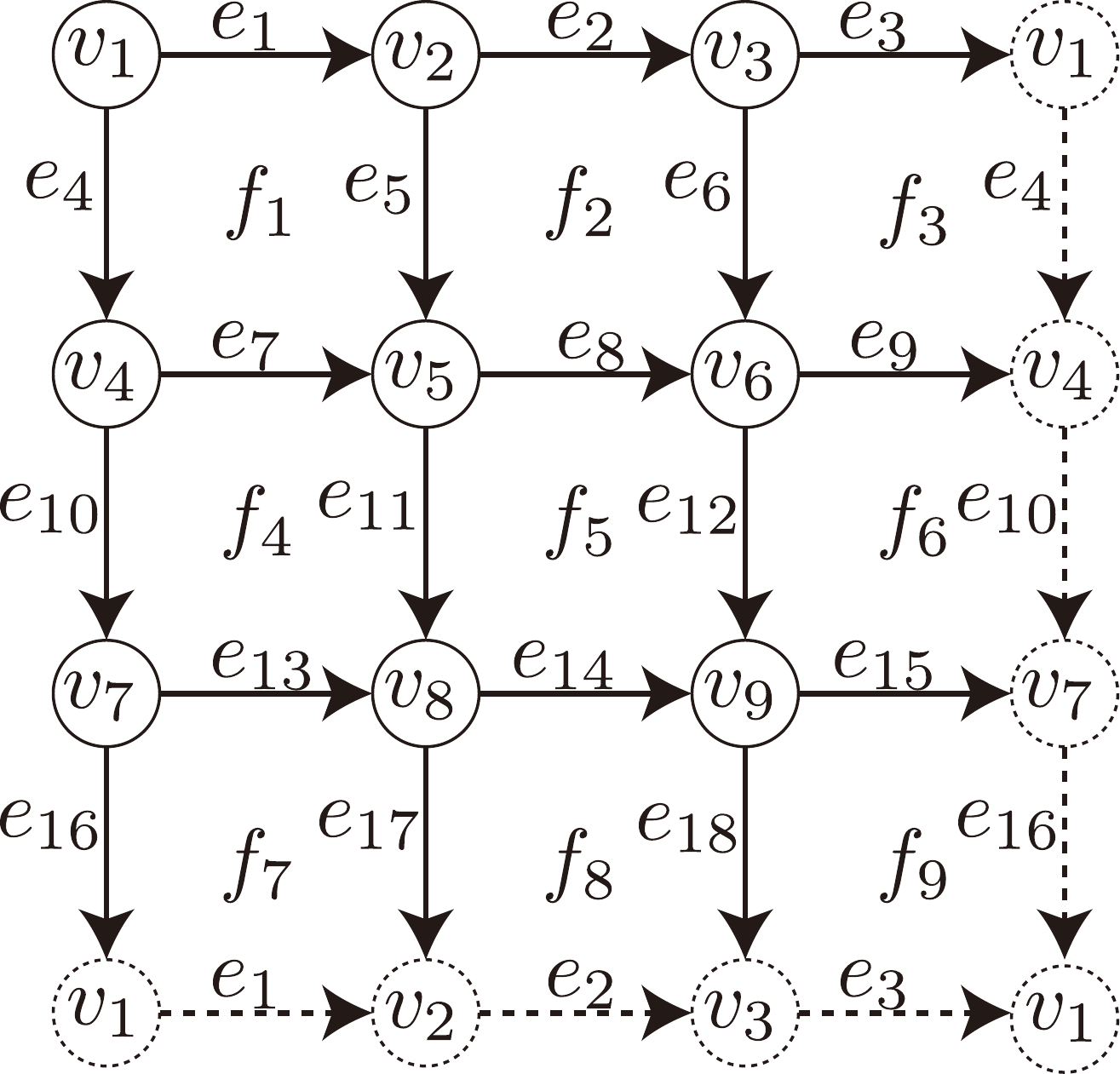}
\end{center}
\caption{A directed graph for a $3\times 3$ torus.
There are 9 vertices, 9 faces
and 18 directed edges.}
\label{torus graph}
\end{figure}

A directed graph for a torus is depicted in Fig.~\ref{torus graph}.
The torus is divided into $3\times 3$ square faces (9 faces in total)
and  the associated graph has a periodicity for two directions. 

We first provide the covariant incidence matrix for this graph. We can immediately reproduce a usual incidence
matrix by setting $U^e=1$ for all.
\be
{\scriptsize
L_U=
\begin{pmatrix}
-1 & U^1\cdot U^{1\dag} & 0 & 0 & 0 & 0 & 0 & 0 & 0 \\
0 & -1 & U^2\cdot U^{2\dag} & 0 & 0 & 0 & 0 & 0 & 0 \\
U^3\cdot U^{3\dag} & 0 & -1 & 0 & 0 & 0 & 0 & 0 & 0 \\
-1 & 0 & 0 & U^4\cdot U^{4\dag}  & 0 & 0 & 0 & 0 & 0 \\
0 & -1 & 0 & 0 & U^5\cdot U^{5\dag} & 0 & 0 & 0 & 0 \\
0 & 0 & -1 & 0 & 0 & U^6\cdot U^{6\dag} & 0 & 0 & 0 \\
0 & 0 & 0 & -1 & U^7\cdot U^{7\dag} & 0 & 0 & 0 & 0 \\
0 & 0 & 0 & 0 & -1 & U^8\cdot U^{8\dag} & 0 & 0 & 0 \\
0 & 0 & 0 & U^9\cdot U^{9\dag} & 0 & -1 & 0 & 0 & 0 \\
0 & 0 & 0 & -1 & 0 & 0 & U^{10}\cdot U^{10\dag} & 0 & 0 \\
0 & 0 & 0 & 0 & -1 & 0 & 0 & U^{11}\cdot U^{11\dag} & 0 \\
0 & 0 & 0 & 0 & 0 & -1 & 0 & 0 & U^{12}\cdot U^{12\dag} \\
0 & 0 & 0 & 0 & 0 & 0 & -1 & U^{13}\cdot U^{13\dag} & 0 \\
0 & 0 & 0 & 0 & 0 & 0 & 0 & -1 & U^{14}\cdot U^{14\dag} \\
0 & 0 & 0 & 0 & 0 & 0 & U^{15}\cdot U^{15\dag} & 0 & -1 \\
U^{16}\cdot U^{16\dag} & 0 & 0 & 0 & 0 & 0 & -1 & 0 & 0 \\
0 & U^{17}\cdot U^{17\dag} & 0 & 0 & 0 & 0 & 0 & -1 & 0 \\
0 & 0 & U^{18}\cdot U^{18\dag} & 0 & 0 & 0 & 0 & 0 & -1
\end{pmatrix}
}.
\ee
The Laplacian matrix on the vertex is
\be
\Delta_V={\footnotesize
\left(
\begin{array}{ccccccccc}
 4 & -1 & -1 & -1 & 0 & 0 & -1 & 0 & 0 \\
 -1 & 4 & -1 & 0 & -1 & 0 & 0 & -1 & 0 \\
 -1 & -1 & 4 & 0 & 0 & -1 & 0 & 0 & -1 \\
 -1 & 0 & 0 & 4 & -1 & -1 & -1 & 0 & 0 \\
 0 & -1 & 0 & -1 & 4 & -1 & 0 & -1 & 0 \\
 0 & 0 & -1 & -1 & -1 & 4 & 0 & 0 & -1 \\
 -1 & 0 & 0 & -1 & 0 & 0 & 4 & -1 & -1 \\
 0 & -1 & 0 & 0 & -1 & 0 & -1 & 4 & -1 \\
 0 & 0 & -1 & 0 & 0 & -1 & -1 & -1 & 4 \\
\end{array}
\right)},
\label{Laplacian on vertex of torus}
\ee
which has the eigenvalues of $\{6,6,6,6,3,3,3,3,0\}$.

The covariant dual incidence matrix is made from the plaquette variables, but
it is a huge size matrix we can not typeset here. To display in the manuscript,
we give the incidence matrix by setting $U^e=1$ ;
\be
\check{L}^T=
{\tiny
\left(
\begin{array}{cccccccccccccccccc}
 -1&0&0&1&-1&0&1&0&0&0&0&0&0&0&0&0&0&0\\
 0&-1&0&0&1&-1&0&1&0&0&0&0&0&0&0&0&0&0\\
 0&0&-1&-1&0&1&0&0&1&0&0&0&0&0&0&0&0&0\\
 0&0&0&0&0&0&-1&0&0&1&-1&0&1&0&0&0&0&0\\
 0&0&0&0&0&0&0&-1&0&0&1&-1&0&1&0&0&0&0\\
 0&0&0&0&0&0&0&0&-1&-1&0&1&0&0&1&0&0&0\\
 1&0&0&0&0&0&0&0&0&0&0&0&-1&0&0&1&-1&0\\
 0&1&0&0&0&0&0&0&0&0&0&0&0&-1&0&0&1&-1\\
 0&0&1&0&0&0&0&0&0&0&0&0&0&0&-1&-1&0&1
\end{array}
\right)}.
\ee
The Laplacian matrix on the face $\Delta_F$ is the same as $\Delta_V$ in (\ref{Laplacian on vertex of torus}).
The Laplacian matrix on the edge is given by
\be
\Delta_E=
{\tiny
\left(
\begin{array}{cccccccccccccccccc}
 4 & -1 & -1 & 0 & 0 & 0 & -1 & 0 & 0 & 0 & 0 & 0 & -1 & 0 & 0 & 0 & 0 & 0 \\
 -1 & 4 & -1 & 0 & 0 & 0 & 0 & -1 & 0 & 0 & 0 & 0 & 0 & -1 & 0 & 0 & 0 & 0 \\
 -1 & -1 & 4 & 0 & 0 & 0 & 0 & 0 & -1 & 0 & 0 & 0 & 0 & 0 & -1 & 0 & 0 & 0 \\
 0 & 0 & 0 & 4 & -1 & -1 & 0 & 0 & 0 & -1 & 0 & 0 & 0 & 0 & 0 & -1 & 0 & 0 \\
 0 & 0 & 0 & -1 & 4 & -1 & 0 & 0 & 0 & 0 & -1 & 0 & 0 & 0 & 0 & 0 & -1 & 0 \\
 0 & 0 & 0 & -1 & -1 & 4 & 0 & 0 & 0 & 0 & 0 & -1 & 0 & 0 & 0 & 0 & 0 & -1 \\
 -1 & 0 & 0 & 0 & 0 & 0 & 4 & -1 & -1 & 0 & 0 & 0 & -1 & 0 & 0 & 0 & 0 & 0 \\
 0 & -1 & 0 & 0 & 0 & 0 & -1 & 4 & -1 & 0 & 0 & 0 & 0 & -1 & 0 & 0 & 0 & 0 \\
 0 & 0 & -1 & 0 & 0 & 0 & -1 & -1 & 4 & 0 & 0 & 0 & 0 & 0 & -1 & 0 & 0 & 0 \\
 0 & 0 & 0 & -1 & 0 & 0 & 0 & 0 & 0 & 4 & -1 & -1 & 0 & 0 & 0 & -1 & 0 & 0 \\
 0 & 0 & 0 & 0 & -1 & 0 & 0 & 0 & 0 & -1 & 4 & -1 & 0 & 0 & 0 & 0 & -1 & 0 \\
 0 & 0 & 0 & 0 & 0 & -1 & 0 & 0 & 0 & -1 & -1 & 4 & 0 & 0 & 0 & 0 & 0 & -1 \\
 -1 & 0 & 0 & 0 & 0 & 0 & -1 & 0 & 0 & 0 & 0 & 0 & 4 & -1 & -1 & 0 & 0 & 0 \\
 0 & -1 & 0 & 0 & 0 & 0 & 0 & -1 & 0 & 0 & 0 & 0 & -1 & 4 & -1 & 0 & 0 & 0 \\
 0 & 0 & -1 & 0 & 0 & 0 & 0 & 0 & -1 & 0 & 0 & 0 & -1 & -1 & 4 & 0 & 0 & 0 \\
 0 & 0 & 0 & -1 & 0 & 0 & 0 & 0 & 0 & -1 & 0 & 0 & 0 & 0 & 0 & 4 & -1 & -1 \\
 0 & 0 & 0 & 0 & -1 & 0 & 0 & 0 & 0 & 0 & -1 & 0 & 0 & 0 & 0 & -1 & 4 & -1 \\
 0 & 0 & 0 & 0 & 0 & -1 & 0 & 0 & 0 & 0 & 0 & -1 & 0 & 0 & 0 & -1 & -1 & 4 \\
\end{array}
\right)},
\ee
which has the eigenvalues $\{6,6,6,6,6,6,6,6,3,3,3,3,3,3,3,3,0,0\}$.

We can also show that the orthogonality $L_U^\dag \check{L}_U=\check{L}^\dag_U L_U=0$ iff $P^f=1$,
explicitly.

\section{Cartan-Weyl basis and properties}
\label{Cartan-Weyl}

We denote the generators in the Cartan subalgebla ${\mathfrak u}(N)$ by 
$\mathtt{H}_i$ ($i=1,\ldots,N$) and
the root vectors (Weyl generators) by $\mathtt{E}_\alpha$.
These generators obey the following commutation relations;
\be
\begin{split}
&[\mathtt{H}_i,\mathtt{H}_j]=0,\\
&[\mathtt{H}_i,\mathtt{E}_{\pm \alpha}]=\pm \alpha_i \mathtt{E}_{\pm\alpha},\\
&[\mathtt{E}_\alpha,\mathtt{E}_{-\alpha}]=\sum_{i=1}^N \alpha_i \mathtt{H}_i,
\qquad [\mathtt{E}_\alpha,\mathtt{E}_\beta] = N_{\alpha,\beta}\mathtt{E}_{\alpha+\beta},
\end{split}
\ee
and has the properties;
\be
\mathtt{E}_\alpha^\dag =\mathtt{E}_{-\alpha},
\quad \Tr \mathtt{E}_\alpha\mathtt{E}_\beta = \delta_{\alpha+\beta,0},
\quad \Tr \mathtt{H}_i \mathtt{H}_j  = \sum_\alpha \alpha_i \alpha_j= \delta_{ij}.
\ee

Any adjoint representation of $U(N)$ group (generators of the Lie algebra) can be expanded by
\be
X = \sum_{i=1}^N x^i \mathtt{H}_i + \sum_{\alpha} x^\alpha \mathtt{E}_\alpha.
\ee

We use these conventions in the manuscript.

\bibliographystyle{JHEP.bst}
\bibliography{refs}

\end{document}